\documentclass[conference]{IEEEtran}
\usepackage{epsfig}
\usepackage{times}
\usepackage{float}
\usepackage{afterpage}
\usepackage{amsmath}
\usepackage{amstext}
\usepackage{amssymb,bm}
\usepackage{latexsym}
\usepackage{color}
\usepackage{graphicx}
\usepackage{amsmath}
\usepackage{amsthm}
\usepackage{graphicx}
\usepackage[center]{caption}
\usepackage{pstricks}
\usepackage{subfigure}
\usepackage{booktabs}
\usepackage{multicol}
\usepackage{lipsum}
\usepackage{dblfloatfix}
\allowdisplaybreaks

\newtheorem{thm}{Theorem}

\newtheorem{lem}[thm]{Lemma}

\newtheorem{defin}{Definition}

\begin{document}

	\title{On Secure Network Coding for Two Unicast Sessions}

	\author{
		\IEEEauthorblockN{Gaurav Kumar Agarwal, Martina Cardone, Christina Fragouli }
		Department of Electrical Engineering \\
		University of California Los Angeles, Los Angeles, CA 90095, USA\\
		Email: \{gauravagarwal, martina.cardone, christina.fragouli\}@ucla.edu
		\thanks{The work of the authors was partially funded by NSF under
			award 1321120. 
			G. K. Agarwal is also supported by the Guru Krupa Fellowship.}}
	\IEEEoverridecommandlockouts
	
	\maketitle
	\begin{abstract}
		This paper characterizes the secret message capacity of three networks where two unicast sessions share some of the communication resources.
		Each network consists of erasure channels with state feedback.
		A passive eavesdropper is assumed to wiretap any one of the links.
		The capacity achieving schemes as well as the outer bounds are formulated as linear programs.
		The proposed strategies are then numerically evaluated and shown to achieve higher rate performances (up to a double single- or sum-rate) with respect to alternative strategies, where the network resources are time-shared among the two sessions.
		These results represent a step towards the secure capacity characterization for general networks.
		{They also show that, even in configurations for which network coding does not offer benefits in absence of security, it can become beneficial under security constraints.}
	\end{abstract}

	\section{Introduction}
	Secure network coding has well established the benefits of network coding for secure multicast transmission.
	We are here interested in a different type of traffic where we have two independent unicast sessions and
	we seek to answer the following question: what are the benefits that `network coding' type operations offer?

\begin{figure}
		\centering
		\includegraphics[width=0.8\columnwidth]{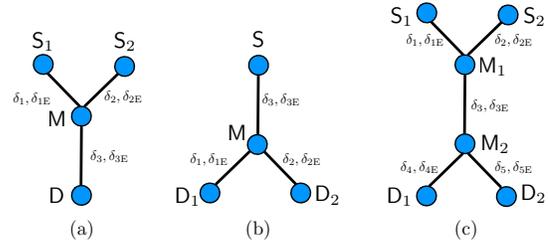}
		\caption{The Y-network in (a), the RY-network in (b) and the X-network in (c).}
		\label{fig:networks}
		\vspace{-5mm}
	\end{figure}

	We consider the three networks in Fig.~\ref{fig:networks}, namely the Y-network, the Reverse Y (RY)-network and the X-network. 
	In the Y-network two sources (able to generate randomness at infinite rate) wish to communicate two independent messages to a common destination, via an intermediate node (unable to generate randomness).
	In the RY-network one source (able to generate randomness at finite rate) aims to communicate two independent messages to two different receivers, through an intermediate node (unable to generate randomness).
	Finally, in the X-network two sources (able to generate randomness at infinite rate) seek to communicate two independent messages to two different receivers, via two intermediate nodes (unable to generate randomness).
	In our network model, the transmissions take place over orthogonal erasure channels; although this being a simplistic assumption,
	yet it captures some intrinsic properties of the wireless medium (such as its lossy nature).
	A passive eavesdropper wiretaps any one of the communication links, but the information about which one is not available\footnote{This assumption is equivalent to have one eavesdropper on every link, but these eavesdroppers do not cooperate among themselves.}.
	Public feedback, which in~\cite{Maurer} was shown to increase the secrecy capacity, is used, i.e., each of the legitimate nodes involved in the communication sends an acknowledgment after each transmission; this is received by {all nodes in the network as well as by the eavesdropper.}
	
	We characterize the secret message capacity for the three networks in Fig.~\ref{fig:networks}.
	The {capacity-achieving} schemes involve two phases: the key-sharing phase and the message-transmission phase. In particular, first a secret key is created between two consecutive legitimate nodes (link-by-link key generation); then, these keys are used to encrypt and transmit the message like in the one-time pad~\cite{ShannonOneTime}. 
	For each of the analyzed networks, the capacity is given as the solution of a Linear Program (LP). 
	We also show, through numerical simulations, the benefits of our schemes compared to two alternative strategies where (a) the two sessions are time-shared and (b) the shared link is time-shared among the two sessions. 
	We prove that `network coding' type operations are beneficial for the three networks in Fig.~\ref{fig:networks}.
	This is because random packets transmitted by different sources can be mixed to create the key to be used on the shared link.
	Similarly, the same set of random packets can be used to generate secret keys for different destinations.
	This result is surprising since, in absence of security considerations, network coding is not beneficial for the networks in Fig.~\ref{fig:networks}.
	\\
	{\bf{Related Work.}}
	The characterization of the secret capacity for wireless networks is a long-standing open problem.
	Relevant work includes~\cite{Wyner}, where the author derived the secret message capacity of the wiretap channel without feedback, and~\cite{Yeung}, where it was shown that secure network coding is optimal for wireless networks with error-free and unit capacity channels.
	The work presented in this paper follows a line of research which was pioneered by the authors in~\cite{czapP2P}, where the secret capacity of the point-to-point channel was characterized and expressed as the solution of an LP. In particular, the {capacity-achieving} scheme proposed by the authors consists of two phases, namely the key-sharing and the message-transmission phases. 
	The same authors extended this approach to characterize the secret capacity of more complex networks, for example: 
	(i) the parallel channel network where a source seeks to securely communicate a message to a destination through a number of independent parallel channels~\cite{CzapVNet}, and
	(ii) the V-network where two sources, which share a common randomness, aim to convey  the same message to a common destination~\cite{CzapVNet}.
	By using a similar approach, in~\cite{PapadopoulosLine} the authors derived the secret message capacity of the line network when the eavesdropper wiretaps one channel as well as all the channels. 
	Recently, in~\cite{PapadopoulosISIT2015} the authors considered a general network and designed two polynomial-time secure transmission schemes. 
	Although the two schemes were not proven 
	{to be capacity-achieving,}
	the work in~\cite{PapadopoulosISIT2015} represents an attempt to characterize the secrecy capacity of an arbitrary network.
	Contrary to all these works, where a single unicast session was considered, here we study the case where two unicast sessions take place simultaneously and share part of the resources. 

	\noindent{\bf{Paper Organization.}}
	Section~\ref{sec:SysModMainResu} describes the three networks of interest, namely the Y-, the RY- and the X-network.
	Section~\ref{sec:ProofCap} presents our main results, i.e., it characterizes the secret message capacity and it presents comparisons with alternative strategies.
	Finally, Section~\ref{sec:Concl} concludes the paper.

	\section{System Model and Main Results}
	\label{sec:SysModMainResu}
	\noindent{\bf{Notation.}} With $\left [ n_1:n_2\right ]$ we denote the set of integers from $n_1$ to $n_2 \geq n_1$. For an index set $\mathcal{A}$ we let $Y_{\mathcal{A}} = \left \{ Y_j: j \in \mathcal{A}\right \}$; $\mathcal{A} \backslash \mathcal{B}$ is the set of elements that belong to $\mathcal{A}$ but not to $\mathcal{B}$. $Y^i$ is a vector of length $i$ with components $\left( Y_1, \ldots, Y_i\right)$.
	
	We consider the three networks in Fig.~\ref{fig:networks}, namely (i) the Y-network in Fig.~\ref{fig:networks}(a) where two sources $\mathsf{S}_1$ and $\mathsf{S}_2$ aim to communicate two independent messages $W_1$ and $W_2$ to a common destination $\mathsf{D}$; (ii) the RY-network in Fig.~\ref{fig:networks}(b) where one source $\mathsf{S}$ has a message $W_1$ for destination $\mathsf{D}_1$ and a message $W_2$ for destination $\mathsf{D}_2$; both in the Y- and in the RY-network the communication occurs through an intermediate/relay node $\mathsf{M}$; finally, we study (iii) the X-network in Fig.~\ref{fig:networks}(c) where two sources $\mathsf{S}_1$ and $\mathsf{S_2}$ seek to communicate two independent messages $W_1$ and $W_2$ to two different destinations $\mathsf{D}_1$ and $\mathsf{D}_2$ via two intermediate nodes $\mathsf{M}_1$ and $\mathsf{M}_2$.
	
	{Each communication link} is an independent erasure channel with one legitimate receiver and one possible passive eavesdropper. In each network, the eavesdropper wiretaps one channel, which one exactly is not {known.}
	The erasure probabilities are denoted as $\delta_j$ and $\delta_{j\text{E}}$, with $j \in [1:3]$ for the Y- and the RY-network and $j \in [1:5]$ for the X-network\footnote{Index $j$ enumerates the channel, e.g., for the RY-network in Fig.~\ref{fig:networks}(b) the channels from $\mathsf{S}$ to $\mathsf{M}$, from $\mathsf{M}$ to $\mathsf{D}_1$ and from $\mathsf{M}$ to $\mathsf{D}_2$ are referred to as channels $3$, $1$ and $2$, respectively.}.
	
	The $j$-th channel input at time instant $i$, with $i \in [1:n]$ (where $n$ is the total number of transmissions), is denoted as $X_{ji} \in \mathbb{F}_q^L$ and referred to as a packet. Without loss of generality, in what follows we let $L \log \left( q\right)=1$, i.e., we express the rate in terms of packets. 
	Similarly, $Y_{ji}$ and $Z_{ji}$ denote the outputs at the legitimate receiver and at the passive eavesdropper, respectively, on channel $j$ at time $i$.
	
	For the three scenarios in Fig.~\ref{fig:networks} we assume public state feedback, i.e., each legitimate node sends an acknowledgment whether the packet transmission was successful, which is received by all other nodes as well as by the eavesdropper. 
	We denote with $F_{ji}$ the feedback of the transmission on channel $j$ at time $i$.
	For the X-network in Fig.~\ref{fig:networks}(c) we have\footnote{Although definitions are given only for the X-network, they straightforwardly extend to the Y- and the RY-network.}
	\begin{align*}
		&\Pr \left \{ Y_{[1:5]i},Z_{[1:5]i} \! \left. \right | \! X_{[1:5]i}\right \} \!=\! \prod_{j=1}^5 \Pr \left \{ Y_{ji} \! \left. \right | \! X_{ji}\right \} \Pr \left \{Z_{ji} \! \left. \right |\! X_{ji} \right \} \!,
		\\&
		\Pr \left \{ Y_{ji} \! \left. \right | \! X_{ji}\right \} =
		\left \{
		\begin{array}{ll}
			1-\delta_j, & Y_{ji} = X_{ji}
			\\
			\delta_j, & Y_{ji} = \bot
		\end{array}
		\right.,
		\\&
		\Pr \left \{ Z_{ji} \! \left. \right | \! X_{ji}\right \} =
		\left \{
		\begin{array}{ll}
			1-\delta_{j\text{E}}, & Z_{ji} = X_{ji}
			\\
			\delta_{j\text{E}}, & Z_{ji} = \bot
		\end{array}
		\right.,
	\end{align*}
	where with $\bot$ we denote the symbol of erasure.

	We assume that the intermediate nodes are unable to generate private randomness\footnote{
		The results here presented readily extend to the case when intermediate nodes can generate private randomness at finite rate.
		If intermediate nodes can generate randomness at infinite rate a naive link-by-link time sharing strategy would be capacity-achieving.
	}. 
	For the Y- and the X-network the sources $\mathsf{S}_1$ and $\mathsf{S}_2$ can generate private randomness $\Theta_1$ and $\Theta_2$ at infinite rate. 
	Differently, for the RY-network the source $\mathsf{S}$ can generate private randomness at a finite rate $D_0$.
	The messages $W_1$ and $W_2$ consist of $N_1$ and $N_2$ packets, respectively, and have to be reliably and securely decoded at the legitimate receiver.
	
	\begin{defin}
		For the X-network in Fig.~\ref{fig:networks}(c) a secure coding scheme with parameters $\left(N_1, N_2, n, \epsilon \right )$ consists of $5$ encoding functions $f_{ji}, \   j \in [1:5]$  {for each $i\in [1:n]$} such that
\begin{align*}
X_{ji} = 
\left \{
\begin{array}{ll}
f_{ji} \left( W_j, \Theta_j, F_{\mathcal{A}}^{i-1}  \right) & \text{if} \ j \in [1:2]
\\
f_{ji} \left( Y_{1}^{i-1},Y_{2}^{i-1}, F_{\mathcal{A}}^{i-1}  \right) & \text{if} \ j = 3
\\
f_{ji} \left( Y_{3}^{i-1}, F_{\mathcal{A}}^{i-1}  \right) & \text{if} \ j \in [4:5]
\end{array}
\right.,
\end{align*}
		where $\mathcal{A}=[1:5]$, and of $2$ decoding functions {$\phi_j$} such that $\mathsf{D}_j$, $j \in [1:2]$, can decode the message $W_j$ with high probability, i.e., $\Pr \left \{ \phi_j \left( Y_{j+3}^n\right)\neq W_j \right \} < \epsilon$. 
		Moreover, the messages $W_1$ and $W_2$ have to remain secret from the eavesdropper, i.e.,
		$I \left( W_1,W_2;Z_k^n,F_{\mathcal{A}}^n\right) < \epsilon, \ \forall k \in [1:5]$.
		A non-negative rate pair $\left( R_1,R_2 \right)$ is securely achievable if, for any $\epsilon >0$, there exists a secure coding scheme with parameters $\left(N_1, N_2, n, \epsilon \right )$ such that $R_j < \frac{1}{n} N_j - \epsilon, \ \forall j \in [1:2]$.
	\end{defin}
	The main contribution of this paper is the {characterization} of the secret message capacity region (the largest securely achievable rate pair $\left( R_1,R_2\right)$) for the three networks in Fig.~\ref{fig:networks} as described in the next three theorems.
	
	\begin{thm}
		\label{eq:CapY}
		The secret message capacity region of the Y-network in Fig.~\ref{fig:networks}(a) with unlimited private randomness at the sources $\mathsf{S}_1$ and $\mathsf{S}_2$ and no private randomness at the relay node $\mathsf{M}$, is the feasible region of the following LP,
		\begin{align*}
			\begin{array}{lll}
				& {\rm{max}} & g(R_1,R_2)
				\\ &{\rm{s.t.}} & k_j  \geq  R_j \frac{1-\delta_{j \text{E}}}{1-\delta_j \delta_{j \text{E}}},   \ j \in [1:2] 
				\\ & & k_3  \geq  (R_1+R_2) \frac{1-\delta_{3\text{E}}}{1-\delta_3 \delta_{3\text{E}}} 
				\\ & & \frac{R_j}{1-\delta_j} + \frac{k_j}{(1-\delta_j)\delta_{j \text{E}}}  \leq  1,  \ j \in [1:2] 
				\\ & &  \frac{R_1+R_2}{1-\delta_3} + \frac{k_3}{(1-\delta_3)\delta_{3\text{E}}}  \leq  1 
				\\ & & k_3 \leq  \left ( \frac{k_1}{\delta_{1\text{E}}}+ \frac{k_2}{\delta_{2\text{E}}} \right ) \frac{(1-\delta_3) \delta_{3\text{E}}}{1-\delta_3 \delta_{3\text{E}}} 
				\\ &  & R_i,k_j \geq 0, \ i \in [1:2], \ j \in [1:3],
			\end{array}
		\end{align*}
		where $g(R_1,R_2)$ can be any linear function {of $\left(R_1,R_2 \right)$.}
	\end{thm}
	
	\begin{thm}
		\label{eq:CapInvY}
		The secret message capacity region of the RY-network in Fig.~\ref{fig:networks}(b) with limited private randomness of rate $D_0$ at the source $\mathsf{S}$ and no private randomness at the relay node $\mathsf{M}$, is the feasible region of the following LP,
		\begin{align*}
			\begin{array}{lll}
				& {\rm{max}} & g(R_1,R_2)
				\\ &{\rm{s.t.}} & k_3 + e \frac{(1-\delta_3) \delta_{3 \text{E}}}{1-\delta_3 \delta_{3\text{E}}}  \geq  (R_1+R_2) \frac{1-\delta_{3 \text{E}}}{1-\delta_3 \delta_{3 \text{E}}} 
				\\ & & k_j  \geq  R_j \frac{1-\delta_{j \text{E}}}{1-\delta_j \delta_{j \text{E}}},    \ j \in [1:2]
				\\ & & \frac{R_1+R_2}{1-\delta_3} + \frac{k_3}{(1-\delta_3)\delta_{3 \text{E}}} + \frac{e}{1-\delta_3}  \leq  1  
				\\ & &  	\frac{R_j}{1-\delta_j} + \frac{k_j}{(1-\delta_j)\delta_{j \text{E}}}  \leq  1,  \  j \in [1:2]
				\\ & & k_3  \leq (D_0 - e) \frac{(1-\delta_3) \delta_{3 \text{E}} }{1-\delta_3 \delta_{3 \text{E}}} 
				\\ & & k_j  \leq  (e + \frac{k_3}{\delta_{3 \text{E}}}) \frac{(1-\delta_j) \delta_{j \text{E}}}{1-\delta_j \delta_{j \text{E}}} , \  j \in [1:2]
				%
				%
				\\ & &  R_i, e,k_j  \geq 0, \ i \in [1:2], \ j \in [1:3],
			\end{array}
		\end{align*}
		where $g(R_1,R_2)$ can be any linear function {of $\left(R_1,R_2 \right)$.}
	\end{thm}
	
	\begin{thm}
		\label{eq:CapButterfly}
		The secret message capacity region of the X-network in Fig.~\ref{fig:networks}(c) with unlimited private randomness at the sources $\mathsf{S}_1$ and $\mathsf{S}_2$ and no private randomness at the relay nodes $\mathsf{M}_1$ and $\mathsf{M}_2$, is the feasible region of the following LP,
		\begin{align*}
			\begin{array}{lll}
				& {\rm{max}} & g(R_1,R_2)
				\\ &{\rm{s. t.}} & 
				k_j  \geq  R_j \frac{1-\delta_{j \text{E}}}{1-\delta_j \delta_{j \text{E}}}, \ j \in [1:2]
				\\ & & k_3 + e \frac{(1-\delta_3) \delta_{3 \text{E}}}{1-\delta_3 \delta_{3\text{E}}} \geq  (R_1+R_2) \frac{1-\delta_{3 \text{E}}}{1-\delta_3 \delta_{3 \text{E}}} 
				\\ & & k_j  \geq R_{j-3} \frac{1-\delta_{j \text{E}}}{1-\delta_j \delta_{j \text{E}}} , \ j \in [4:5]
				\\ & &  	\frac{R_j}{1-\delta_j} + \frac{k_j}{(1-\delta_j)\delta_{j \text{E}}}    \leq  1, \  j \in [1:2]
				\\ & &  	\frac{R_{j-3}}{1-\delta_j} + \frac{k_j}{(1-\delta_j)\delta_{j \text{E}}}  \leq  1,  \ j \in [4:5]
				\\ & & \frac{R_1+R_2}{1-\delta_3} + \frac{k_3}{(1-\delta_3)\delta_{3 \text{E}}} + \frac{e}{1-\delta_3}  \leq  1 
				\\ & & k_3  \leq  ( \frac{k_1}{\delta_{1\text{E}}}+  \frac{k_2}{\delta_{2\text{E}}} - e) \frac{(1-\delta_3) \delta_{3 \text{E}}  }{1-\delta_3 \delta_{3 \text{E}}}  
				\\ & & k_j  \leq   (e+ \frac{k_3}{\delta_{3\text{E}}}) \frac{(1-\delta_j) \delta_{j \text{E}}}{1-\delta_j \delta_{j \text{E}}} , \  j \in [4:5]
				%
				%
				\\ & &  R_i, e,k_j \geq 0, \ i \in [1:2], \ j \in [1:5],
			\end{array}
		\end{align*}
		where $g(R_1,R_2)$ can be any linear function {of $\left(R_1,R_2 \right)$.}
	\end{thm}
	
	\section{Secure Capacity Characterization}
	\label{sec:ProofCap}
	{We here describe the secure coding schemes\footnote{
			Schemes consider the expected number of transmissions needed.
			Similar to~\cite{czapP2P}, it can be shown that the number of transmissions needed concentrates exponentially fast around the average enabling to achieve expected rate values.
		} 
		and the outer bounds and formulate them as LPs.} We then show through numerical evaluations the benefits (in terms of achievable secure rate) of our scheme with respect to two naive strategies: (i) {\it path sharing}, {i.e.,} the whole communication resources are time-shared among the two sessions and (ii) {\it link sharing},  {i.e.,} the shared link is time-shared among the two sessions.
	
	
	\subsection{Achievability}
	Our secure coding schemes for the networks in Fig.~\ref{fig:networks} consist of two phases, namely the key-sharing and the message-transmission. In what follows we describe these two phases and explain how these relate to the LPs in Theorems~\ref{eq:CapY}-\ref{eq:CapButterfly}.
	
	\subsubsection{The Y-network}
	On channel $j \in [1:2]$, source $\mathsf{S}_j$ sends $\frac{k_j}{(1-\delta_j)\delta_{j \text{E}}}$ independent random packets generated from her private randomness (assumed to be infinite). 
	Out of these, a total of $k_j$ packets are received by the relay node $\mathsf{M}$, but not by the possible eavesdropper.
	We do not know exactly which packets, out of the $\frac{k_j}{\delta_{j \text{E}}}$ ones received by $\mathsf{M}$, are also received by the possible eavesdropper. However, out of the packets received by $\mathsf{M}$, we can always create $ k_j $ independent packets, which are also independent of the packets received by the possible eavesdropper. 
	We do this by multiplying the $\frac{k_j}{\delta_{j \text{E}}}$ packets of $\mathsf{M}$ by an MDS code matrix of dimension {$\left [ \frac{k_j}{\delta_{j \text{E}}}\right ] \times \left [ k_j \right ]$}. 
	{Thus, without loss of generality, we can assume that we always know which packets are received by the legitimate node and not by the possible eavesdropper, if we know their amount~\cite{PapadopoulosLine}.}
	All these packets are used to generate a {\it secret key} on channel $j \in [1:2]$ between nodes $\mathsf{S}_j$ and $\mathsf{M}$ ({\it key-sharing} phase).
	These packets are then expanded by means of an MDS code matrix of size $\left [ k_j  \right ] \times \left [ R_j \right ] $ and used as in the one-time pad~\cite{ShannonOneTime} to encrypt $R_j$ message packets, which are sent using the ARQ protocol ({\it message-transmission} phase). 
	
	At the intermediate node $\mathsf{M}$ (assumed to be unable to generate any randomness) there are $\left( \frac{k_1}{\delta_{1\text{E}}}\right) + \left(\frac{k_2}{\delta_{2\text{E}}} \right)$ available random packets (received from $\mathsf{S}_1$ and $\mathsf{S}_2$ on channels $1$ and $2$, respectively). 
	By means of an MDS code these random packets are first expanded by a factor $\frac{1}{1-\delta_3 \delta_{3\text{E}}}$ and then only $\frac{k_3}{(1-\delta_3)\delta_{3\text{E}}}$ of them are sent to node $\mathsf{D}$.
	With this, the number of random packets received by $\mathsf{D}$, but not by the possible eavesdropper is $k_3$. These random packets are used to generate a {\it secret key} on channel $3$ between nodes $\mathsf{M}$ and $\mathsf{D}$ ({\it key-sharing} phase).
	Similar to channels $1$ and $2$, also for channel $3$ we expand these $k_3$ packets by means of an MDS code matrix of dimension $\left [k_3 \right ]  \times \left [ R_1+R_2\right ]$ and then we use them to encrypt $R_1+R_2$ message packets as in the one-time pad~\cite{ShannonOneTime}. These message packets are finally transmitted by using the ARQ protocol ({\it message-transmission} phase). 
	
	The scheme described above is equivalent to the LP in Theorem~\ref{eq:CapY}, where the variables $R_i$ and $k_j$ , with $i \in [1 : 2]$ and $j \in [1 : 3]$, represent the message rate for the pair $\mathsf{S}_i-\mathsf{D}$ and the key created on channel $j$, respectively. 
	In particular: (i) the {\it first} and {\it second} inequalities are security constraints, {i.e.,} they ensure that the key that is generated is greater than the key which is consumed\footnote{Since the encrypted packets are sent by using the ARQ protocol, the key consumed on channel $j \in [1:3]$ (i.e., the number of packets received by the possible eavesdropper on that channel) is $R_j \frac{1-\delta_{j \text{E}}}{1-\delta_j \delta_{j \text{E}}}$, with $R_3=R_1+R_2$.}; (ii) the {\it third} and {\it fourth} inequalities are time constraints, i.e., the length of the key generation phase plus the length of the message sending phase 
	cannot exceed the total available time; (iii) the {\it fifth} inequality {follows since} node $\mathsf{M}$ has zero randomness and so the key that it can create is constrained by the randomness {received} from $\mathsf{S}_1$ and $\mathsf{S}_2$.

	\subsubsection{The RY-network}
	{In~\cite{PapadopoulosLine}, the authors showed that, in a line network where a node has limited randomness and the next node can generate randomness based on the one received from the previous node(s), a combination of ARQ and MDS coding is needed for optimally generating the key. 
		Following this, on channel $3$ of the RY-network, the source $\mathsf{S}$ transmits $e$ independent random packets using the ARQ protocol. These packets are all received by the relay node $\mathsf{M}$, while the possible eavesdropper receives a fraction {$ \frac{1- \delta_{3 \text{E}}}{1-\delta_3 \delta_{3\text{E}}}$} of them.}
	By means of an MDS code, the remaining $\left( D_0 - e\right)$ random packets at the source $\mathsf{S}$ are expanded by a factor $\frac{1}{1-\delta_3\delta_{3\text{E}}}$ and then only $\frac{k_3}{(1-\delta_3)\delta_{3\text{E}}}$ of them are sent to node $\mathsf{M}$.
	Thus, the total number of packets received by the intermediate node $\mathsf{M}$, but not by the possible eavesdropper is $k_3 + e \frac{(1-\delta_3) \delta_{3 \text{E}}}{1-\delta_3 \delta_{3\text{E}}}$.
Similar to the Y-network, from the $e+\frac{k_3}{\delta_{3 \text{E}}}$ independent random packets received by $\mathsf{M}$, we generate $k_3 + e \frac{(1-\delta_3) \delta_{3 \text{E}}}{1-\delta_3 \delta_{3\text{E}}}$ packets. 
	All these packets are used to generate a {\it secret key} on channel $3$ between nodes $\mathsf{S}$ and $\mathsf{M}$ ({\it key-sharing} phase).
	They are then expanded by means of an MDS code matrix of size $\left [ k_3 + e \frac{(1-\delta_3) \delta_{3 \text{E}}}{1-\delta_3 \delta_{3\text{E}}} \right ] \times \left [R_1+R_2\right ]$ and used as in the one-time pad~\cite{ShannonOneTime} to encrypt $R_1+R_2$ message packets, which are sent using the ARQ protocol ({\it message-transmission} phase).  
	
	At the relay node $\mathsf{M}$ (assumed unable to generate any randomness) there are $\left( \frac{k_3}{\delta_{3\text{E}}} + e\right)$ available random packets (received from $\mathsf{S}$ on channel $3$). 
	By means of an MDS code these random packets $\forall j \in [1:2]$ are first expanded by a factor $\frac{1}{1-\delta_j\delta_{j\text{E}}}$ and then only $\frac{k_j}{(1-\delta_j)\delta_{j\text{E}}}$ of them are sent to node $\mathsf{D}_j$ on channel $j$.
	With this, the number of random packets received by $\mathsf{D}_j$, but not by the possible eavesdropper is $k_j$. These packets are used to generate a {\it secret key} on channel $j$ between nodes $\mathsf{M}$ and $\mathsf{D}_j$ ({\it key-sharing} phase).
	These $k_j$ packets are then expanded by means of an MDS code matrix of dimension $\left [k_j \right ] \!\times\! \left [R_j \right ]$ and used to encrypt $R_j$ message packets as in the one-time pad~\cite{ShannonOneTime}. These message packets are finally transmitted using ARQ ({\it message-transmission} phase).
	
	{Similar to} the Y-network, also for the RY-network the secure transmission strategy above is equivalent to the LP in Theorem~\ref{eq:CapInvY}.
	The variables $R_i$, $k_j$ and $e$, with $i \in [1 : 2]$ and $j \in [1 : 3]$, represent the message rate for the pair $\mathsf{S}-\mathsf{D}_i$, the key created on channel $j$, and the extra randomness (in addition to the one sent for generating the key $k_3$) sent from the source $\mathsf{S}$, respectively. 
	In particular: (i) the {\it first} and {\it second} inequalities are security constraints; (ii) the {\it third} and the {\it fourth} inequalities are time constraints; (iii) the {\it fifth} (respectively, {\it sixth}) inequality is due to the fact that the key that node $\mathsf{S}$ (respectively, $\mathsf{M}$) can create is constrained by its limited randomness (respectively, the randomness that it gets from $\mathsf{S}$).
	\subsubsection{The X-network}
	The secure transmission strategy here proposed for the X-network in Fig.~\ref{fig:networks}(c) consists of a mix of the two schemes designed for the Y- and the RY-network. 
	In particular, on channels $1$ and $2$ we use exactly the same operations used on channels $1$ and $2$ of the Y-network, while on channels $3$-$5$ the same strategy proposed for the RY-network applies, with the small difference that the available finite randomness at node $\mathsf{S}$ of the RY-network (node $\mathsf{M}_1$ in the X-network) is now replaced by $D_0= \left( \frac{k_1}{\delta_{1\text{E}}} \right) + \left(\frac{k_2}{\delta_{2\text{E}}} \right)$.
	\subsection{Converse}
	
	We here highlight the main steps to derive an outer bound on the secure capacity for the networks in Fig.~\ref{fig:networks} and to formulate it as an LP; the complete proof can be found in the Appendix.
	\\
	{\bf{Step 1.}} We prove and make use of the following lemma (see Appendix \ref{app:proofLemma} for the details), which is a generalization of those in~\cite{PapadopoulosLine} for the line network.
	\begin{lem}
		\label{lemma:lemma1}
		For any $j \in \mathcal{A}$, we have
		\begin{subequations}
			\begin{align}
				&\left( 1- \delta_j\right) \delta_{j\text{E}} \sum_{i=1}^n H \left( X_{ji} \left | \right. Y_j^{i-1}, Z_j^{i-1}, F_j^{i-1},W_{\mathcal{B}},F_{\mathcal{A} \backslash j}^n\right) -\nonumber
				\\&
				\left( 1-\delta_{j\text{E}} \right) \sum_{i=1}^n I \left(Y_j^{i-1},F_j^{i-1}; X_{ji} \left | \right. Z_{j}^{i-1},F_j^{i-1},W_{\mathcal{B}},F_{\mathcal{A} \backslash j}^n \right) \nonumber
				\\&
				= H \left( Y_j^n \left | \right. W_{\mathcal{B}}, F_{\mathcal{A}}^n, Z_j^n \right), \label{eq:firsteqLemma}
				\\&\left( 1- \delta_j\right)  \sum_{i=1}^n H \left( X_{ji} \left | \right. Y_j^{i-1}, F_j^{i-1},W_{\mathcal{B}},F_{\mathcal{A} \backslash j}^n\right) \nonumber
				\\&
				= H \left( Y_j^n \left | \right. W_{\mathcal{B}}, F_{\mathcal{A}}^n \right), \label{eq:extraLemma}
				\\& \sum_{i=1}^n I \left( W_{\mathcal{B}}; X_{ji} \left | \right. F_{\mathcal{A} \backslash j}^n, Z_{j}^{i-1},F_j^{i-1} \right) < \frac{\epsilon}{ 1- \delta_{j\text{E}}},  \label{eq:SeceqLemma}
				\\&
				\sum_{i=1}^n I \left( W_{\mathcal{B}};X_{ji} \left | \right. Y_j^{i\!-\!1},Z_j^{i\!-\!1},F_j^{i\!-\!1},F_{\mathcal{A} \backslash j}^n \right)
				\!\geq \!\frac{n R_j}{1\!-\!\delta_{j\text{E}} \delta_j},  \label{eq:ThirdeqLemma}
			\end{align}
			where: (i) for the Y-network $\mathcal{A}=[1:3]$, $\mathcal{B}=\{j\}$, $W_3 =  W_{[1:2]}$ and $R_3 = R_1+R_2$; (ii) for the RY-network $\mathcal{A}=[1:3]$, $\mathcal{B}=[1:2]$ and $R_3 = R_1+R_2$; (iii) for the X-network $\mathcal{A}=[1:5]$, $\mathcal{B}=\{j\}$, $W_3=W_4=W_5 =  W_{[1:2]}$, $R_3 = R_1+R_2$, $R_4=R_1$ and $R_5=R_2$.
			Moreover, for node $\mathsf{D}$ in the Y-network we have
			\begin{align}
				n R_3 \!\leq\! \left( 1\!-\!\delta_3 \right) \sum_{i=1}^n\! I \left( W_1,W_2;X_{3i} \left | \right. F_{\mathcal{A} \backslash 3}^n,Y_3^{i\!-\!1},F_3^{i\!-\!1}\right)\!, \label{eq:FourtheqLemma}
			\end{align}
			for node $\mathsf{D}_j$, $j \in [1:2]$ in the RY-network, we have 
			\begin{align}
				n  R_j \!\leq\! \left( 1\!-\!\delta_j \right) \sum_{i=1}^n I \left( W_j;X_{ji} \left | \right. F_{\mathcal{A} \backslash j}^n,Y_j^{i\!-\!1},F_j^{i\!-\!1}\right)\!, \label{eq:FiftheqLemma}
			\end{align}
			and for node $\mathsf{D}_j$, $j \in [1:2]$ in the X-network, we have 
			\begin{align}
				n  R_j \!\leq\! \left( 1\!-\!\delta_{j\!+\!3} \right)\! \sum_{i=1}^n \!I \left( W_j;X_{j\!+\!3i} \left | \right. \! F_{\mathcal{A} \backslash j+3}^n,Y_{j\!+\!3}^{i\!-\!1},F_{j\!+\!3}^{i\!-\!1}\right)\!. \label{eq:SixtheqLemma}
			\end{align}
		\end{subequations}
	\end{lem}
	\begin{figure*}
		\begin{subequations}
			\label{eq:corrToT}
			\begin{align}
				&  n k_j \leftrightarrow \frac{\delta_{j\text{E}} (1-\delta_j)}{\delta_j (1-\delta_{j\text{E}})}  \left( \left( 1- \delta_j \delta_{j\text{E}} \right) \sum_{i=1}^n H \left( X_{ji} \left | \right. Y_j^{i-1}, Z_j^{i-1}, F_j^{i-1},W_{\mathcal{B}},F_{\mathcal{A} \backslash j}^n\right) - H \left( Y_j^n | W_{\mathcal{B}},F_{\mathcal{A} }^n \right) \right) ,
				\label{eq:corr1}
				\\ & \ n e_j \leftrightarrow \frac{1-\delta_j\delta_{j\text{E}}}{\delta_j (1-\delta_{j\text{E}})} \left( H \left( Y_j^n | W_{\mathcal{B}},F_{\mathcal{A} }^n \right) - \left( 1-\delta_j\right) \sum_{i=1}^n H \left( X_{ji}\left | \right. Y_j^{i-1},Z_j^{i-1},F_{\mathcal{A}\backslash j }^n,F_j^{i-1},W_{\mathcal{B}} \right) \right),
				\label{eq:corr2}
				\\& n k_{\kappa} \leftrightarrow   \delta_{\kappa \text{E}}  \left( 1- \delta_{\kappa} \right) \sum_{i=1}^n H \left( X_{\kappa i} \left | \right. Y_{\kappa}^{i-1}, Z_{\kappa}^{i-1}, F_{\kappa}^{i-1},W_1,W_2, F_{\mathcal{A} \backslash \kappa}^n\right),
				\label{eq:corr3}
				\\& n k_{\lambda} \leftrightarrow   \delta_{\lambda \text{E}}  \left( 1- \delta_{\lambda} \right) \sum_{i=1}^n H \left( X_{\lambda i} \left | \right. Y_{\lambda}^{i-1}, F_{\lambda}^{i-1},W_\lambda, F_{\mathcal{A} \backslash \lambda}^n\right).
				\label{eq:corr4}
			\end{align}		
		\end{subequations}
		\vspace{-1mm}
		\hrulefill
		\\
		\centering
		\subfigure[Y-network: 
		$\left( \delta_1,\delta_{1\text{E}}\right)\!=\!\left(0.2,0.05 \right)$, 
		$\left( \delta_2,\delta_{2\text{E}}\right)\!=\!\left(0.3,0.05 \right)$,
		$\left( \delta_3,\delta_{3\text{E}}\right)\!=\!\left(0.25,0.05 \right)$.]{
			\includegraphics[width=0.63\columnwidth]{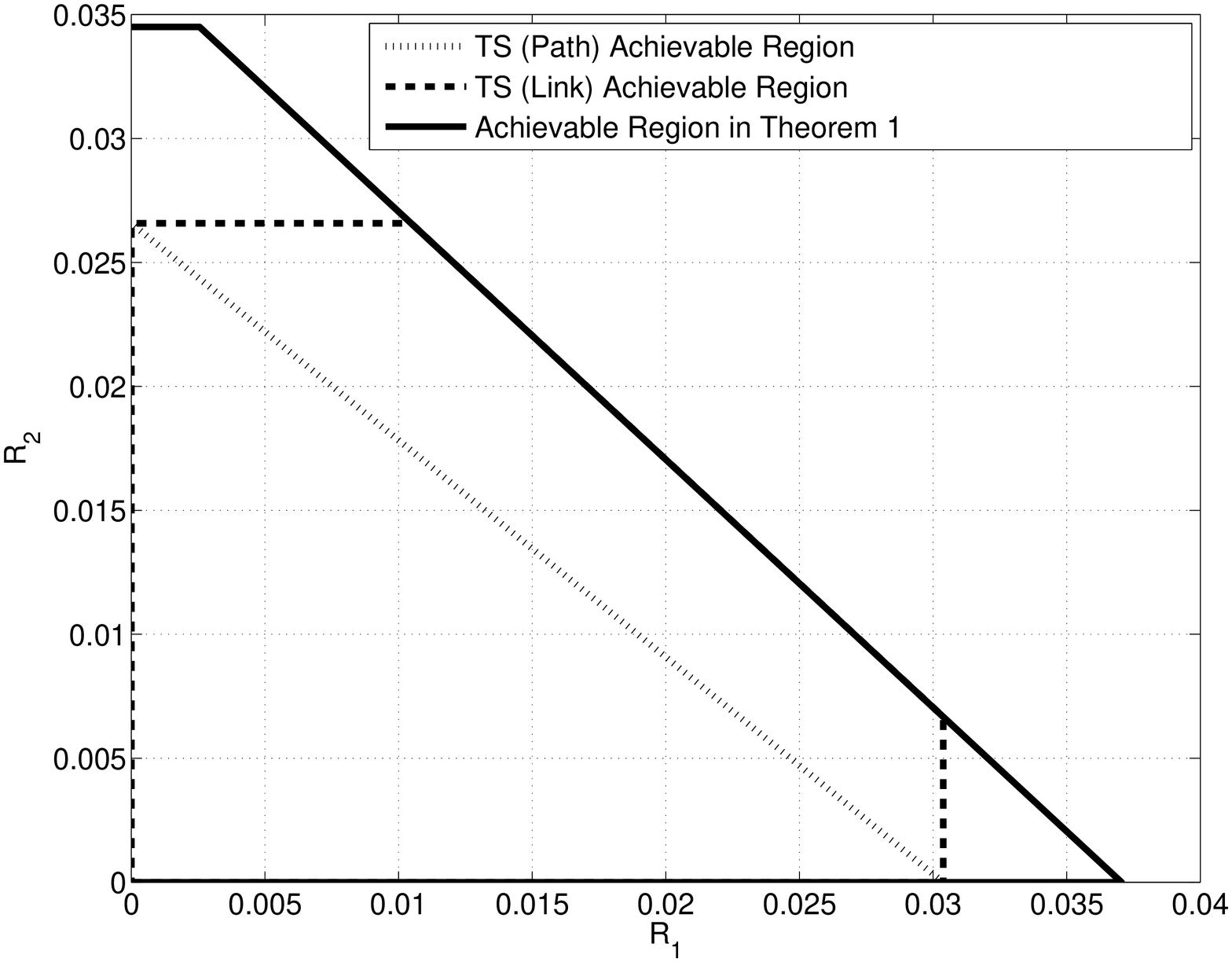}
			\label{fig:YnSim}
		}
		\hfill
		\subfigure[RY-network: 
		$\left( \delta_1,\delta_{1\text{E}}\right)\!=\!\left(0.1,0.1 \right)$, 
		$\left( \delta_2,\delta_{2\text{E}}\right)\!=\!\left(0.2,0.05 \right)$,
		$\left( \delta_3,\delta_{3\text{E}}\right)\!=\!\left(0.3,0.15 \right)$, $D_0\!=\!0.4$.]{
			\includegraphics[width=0.63\columnwidth]{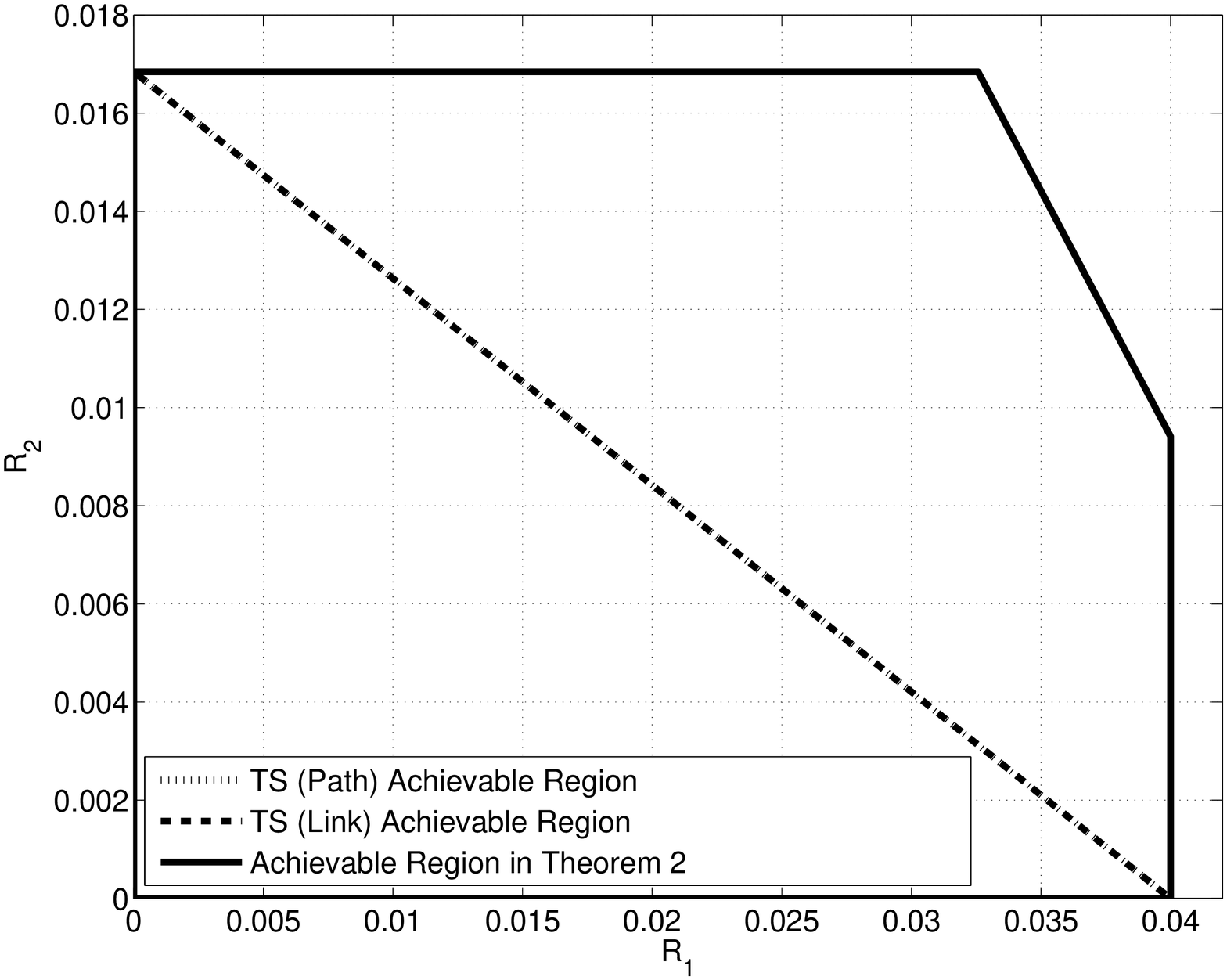}%
			\label{fig:RYnSim}
		}
		\hfill
		\subfigure[X-network:
		$\left( \delta_1,\delta_{1\text{E}}\right)\!=\!\left(0.1,0.1 \right)$, 
		$\left( \delta_2,\delta_{2\text{E}}\right)\!=\!\left(0.2,0.05 \right)$,
		$\left( \delta_3,\delta_{3\text{E}}\right)\!=\!\left(0.3,0.15 \right)$, 
		$\left( \delta_4,\delta_{4\text{E}}\right)\!=\!\left(0.4,0.35 \right)$,
		$\left( \delta_5,\delta_{5\text{E}}\right)\!=\!\left(0.5,0.2 \right)$.
		]{
			\includegraphics[width=0.63\columnwidth]{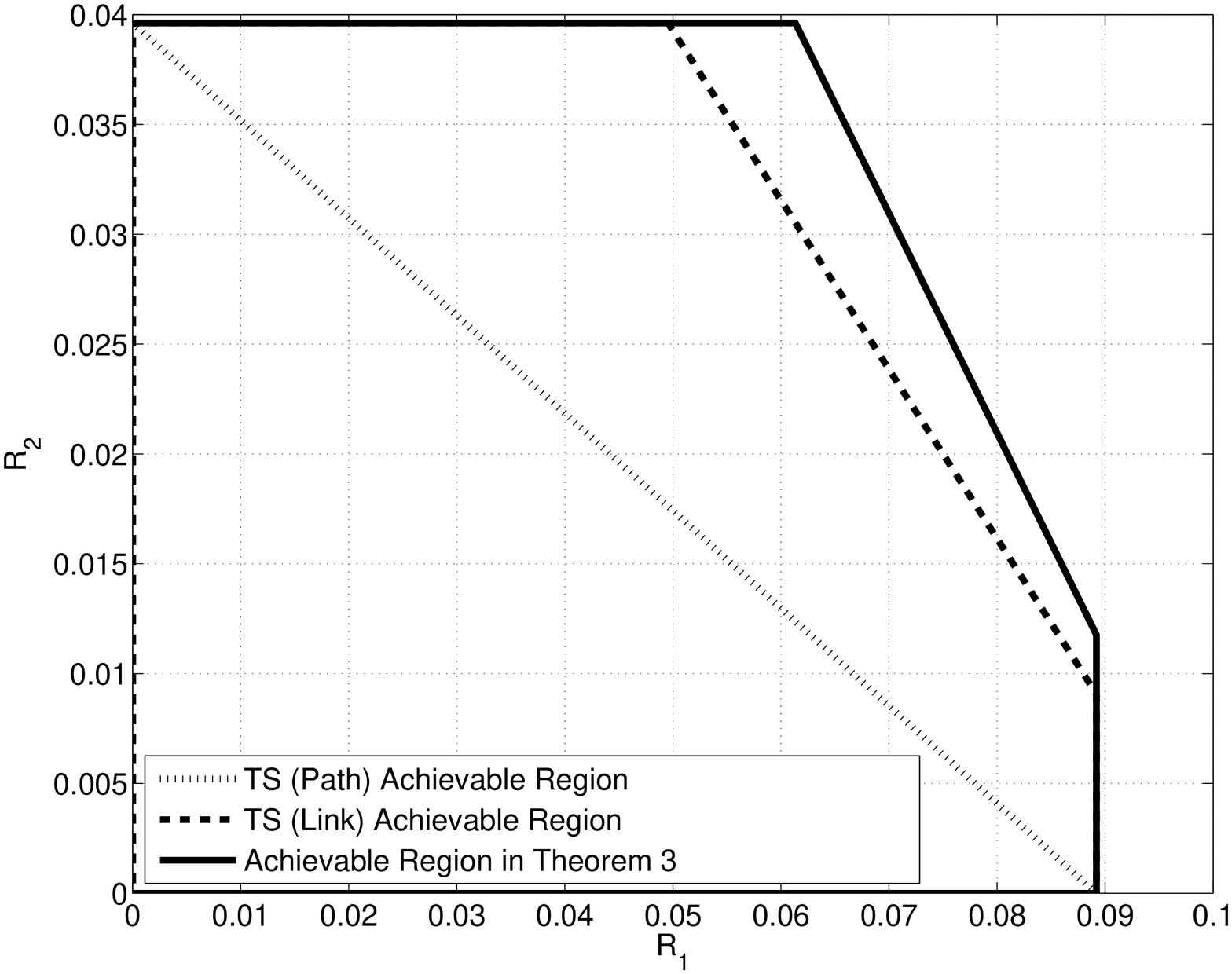}
			\label{fig:XnSim}
		}
		\vspace{-3mm}
		\caption{Numerical evaluations for the three networks in Fig.~\ref{fig:networks}.}
		\label{fig:NumEv}
		\vspace{-4mm}
	\end{figure*}
	\noindent {\bf{Step 2.}} We use the correspondences between {the terms} in~\eqref{eq:corrToT} at the top of the next page {with}: (i) $\mathcal{A}\!=\![1:3]$, $\kappa \! = \! 3$, $\lambda \! \in \! [1:2]$ and $W_3\! =\!  W_{[1:2]}$  for the Y-network; 
	(ii) $\mathcal{A}\!=\![1:3]$, $j \!=\! 3$, {$\kappa \!\in\! [1:2]$}, $\mathcal{B} \!=\! [1:2]$, $W_3\! =\!  W_{[1:2]}$ and $e_3\!=\!e$ for the RY-network; 
	(iii) $\mathcal{A}\!=\![1:5]$, $j \!=\! 3$, $\lambda \! \in \! [1:2]$, $\kappa \! \in \! [4:5]$, $\mathcal{B} \!=\! \{j\}$, $W_3\!=\!W_4\!=\!W_5\! =\!  W_{[1:2]}$ {and $e_3\!=\!e$} for the X-network. 
	{All the quantities} in~\eqref{eq:corrToT} are non-negative (see Appendix B-D for the details).
	\\
	{\bf{Step 3.}} 
	By using the correspondences in~\eqref{eq:corrToT}, Lemma~\ref{lemma:lemma1} and information theoretic properties we derive an outer bound on the secure capacity (see Appendices B-D for the details). 
	Each constraint in the LPs in Theorems~\ref{eq:CapY}-\ref{eq:CapButterfly} is proved to match an outer bound.
	\subsection{Numerical Evaluations}
	We here compare the secrecy capacity performance of our schemes in Theorems~\ref{eq:CapY}-\ref{eq:CapButterfly} with respect to two naive strategies, {i.e.,} the {\it path sharing} and the {\it link sharing}.
	In the {\it path sharing} the whole communication resources, at each time instant, are used only by {one session;} 
	for example, for the X-network
	{we have a time-sharing between $\mathsf{S}_1$-$\mathsf{M}_1$-$\mathsf{M}_2$-$\mathsf{D}_1$ and $\mathsf{S}_2$-$\mathsf{M}_1$-$\mathsf{M}_2$-$\mathsf{D}_2$.}
	Differently, in the {\it link sharing} strategy only the shared communication link is time-shared among the two unicast sessions; for example, in the X-network only the {$\mathsf{M}_1$-$\mathsf{M}_2$ link is time-shared. }
	{
		For both these strategies we do not allow the source node that does not participate to act as a source of randomness, e.g., for the X-network the random packets sent by $\mathsf{S}_1$ cannot be used to encrypt the message packets of $\mathsf{S}_2$.
	}
	Fig.~\ref{fig:NumEv} shows the performance (in terms of secrecy capacity region) of these two time-sharing strategies and of our schemes in Theorems~\ref{eq:CapY}-\ref{eq:CapButterfly}. 
	From Fig.~\ref{fig:NumEv}  we observe that our schemes in Theorems~\ref{eq:CapY}-\ref{eq:CapButterfly} (solid line) achieve higher rates compared to the two time-sharing strategies. 
	Large rate gains are attained when, for each channel, the eavesdropper receives almost {everything and} the legitimate node receives almost no information. 
	{Under these channel} conditions, for the Y-network the individual rates are double than those achieved by the link-sharing strategy; for the RY- and X-network the sum-rate is twice than {that} of the link-sharing scheme.
	In general, these gains follow since: 
	(i) in the Y-network {$\mathsf{S}_1$ and $\mathsf{S}_2$} transmit random packets to $\mathsf{M}$ and {these can be mixed} to create a key on the shared link; 
	(ii) in the RY-network the same set of  random packets can be used to generate keys for both the $\mathsf{M}$-$\mathsf{D}_1$ and $\mathsf{M}$-$\mathsf{D}_2$ links. 
	These factors decrease the number of random packets required to be sent from the source(s) and implies that more message packets can be carried.
	Finally, (iii) in the X-network we have the benefits of both the Y- and RY-network.


	\section{Conclusions}
	\label{sec:Concl}
	We characterized the secret capacity for networks where two unicast sessions share one communication link. 
	This was attained by designing schemes and by deriving outer bounds which were formulated as LPs.
	Through numerical evaluations we showed that our transmission strategies achieve higher rates compared to schemes where the communication resources are time-shared among the two sessions. 
	{These results show that, even in network configurations for which network coding does not offer benefits in absence of security, it can become beneficial under security constraints.}

\appendices

\section{Proof of Lemma \ref{lemma:lemma1}}
\label{app:proofLemma}
We here prove the result in Lemma \ref{lemma:lemma1}.
We start by analyzing the Y-network. 
We have, $\forall j \in \mathcal{A}= [1:3]$ and with $W_3= W_{[1:2]}$,
\begin{align*}
& H \left( Y_j^n | W_j,F_{\mathcal{A} }^n, Z_j^n \right) 
\\& = H \left( Y_j^n, F_j^n | W_j,F_{\mathcal{A} }^n, Z_j^n \right) 
\\& \stackrel{{\rm{(a)}}}{=}  H \left( Y_j^{n-1}, F_j^{n-1} | W_j,F_{\mathcal{A} }^n, Z_j^n \right) 
\\& +  H \left( Y_{jn}, F_{jn} | W_j,F_{\mathcal{A} }^n, Z_j^n, Y_j^{n-1} \right)
\\& \stackrel{{\rm{(b)}}}{=}  H \left( Y_j^{n-1}, F_j^{n-1} | W_j,F_{\mathcal{A}\backslash j }^n, Z_j^{n-1} , F_j^{n-1}\right) 
\\& - I \left( Y_j^{n-1}, F_j^{n-1} ; Z_{jn}, F_{jn} | W_j,F_{\mathcal{A}\backslash j }^n, Z_j^{n-1} , F_j^{n-1} \right)
\\& +  H \left( Y_{jn} | W_j,F_{\mathcal{A} }^n, Z_j^n, Y_j^{n-1} \right)
\\& \stackrel{{\rm{(c)}}}{=}  H \left( Y_j^{n-1}, F_j^{n-1} | W_j,F_{\mathcal{A}\backslash j }^n, Z_j^{n-1} , F_j^{n-1}\right) 
\\&- I \left( Y_j^{n-1}, F_j^{n-1} ; Z_{jn} | W_j,F_{\mathcal{A}\backslash j }^n, Z_j^{n-1} , F_j^{n-1}, F_{jn} \right)
\\& +  H \left( Y_{jn} | W_j,F_{\mathcal{A} }^n, Z_j^n, Y_j^{n-1} \right)
%
%
\\& =  H \left( Y_j^{n-1}, F_j^{n-1} | W_j,F_{\mathcal{A}\backslash j }^n, Z_j^{n-1} , F_j^{n-1}\right) 
\\& - \left( 1- \delta_{jE}\right) I \left( Y_j^{n-1}, F_j^{n-1} ; X_{jn}| W_j,F_{\mathcal{A}\backslash j }^n, Z_j^{n-1} , F_j^{n-1}\right) 
\\& +  \left( 1- \delta_{j}\right) \delta_{jE} H \left( X_{jn} | W_j,F_{\mathcal{A}\backslash j }^n,  F_j^{n-1}, Z_j^{n-1}, Y_j^{n-1} \right) ,
\end{align*}
where:
(i) the equality in ${\rm{(a)}}$ follows from the chain rule of the entropy;
(ii) the equality in ${\rm{(b)}}$ is due to the definition of mutual information;
(iii) finally, the equality in ${\rm{(c)}}$ is because $F_{jn}$ is independent of the rest of the random variables.
By recursively proceeding in the same way for $H \left( Y_j^{n-1}, F_j^{n-1} | W_j,F_{\mathcal{A}\backslash j }^n, Z_j^{n-1} , F_j^{n-1}\right) $, we get the result in \eqref{eq:firsteqLemma}. We now prove \eqref{eq:extraLemma} which is similar to \eqref{eq:firsteqLemma}.

We have, $\forall j \in \mathcal{A}= [1:3]$ and with $W_3= W_{[1:2]}$,
\begin{align*}
& H \left( Y_j^n | W_j,F_{\mathcal{A} }^n\right) 
\\& = H \left( Y_j^n, F_j^n | W_j,F_{\mathcal{A} }^n \right) 
\\& \stackrel{{\rm{(a)}}}{=}  H \left( Y_j^{n-1}, F_j^{n-1} | W_j,F_{\mathcal{A} }^n \right) 
\\& +  H \left( Y_{jn}, F_{jn} | W_j,F_{\mathcal{A} }^n, Y_j^{n-1} \right)
\\& \stackrel{{\rm{(b)}}}{=}  H \left( Y_j^{n-1}, F_j^{n-1} | W_j,F_{\mathcal{A}\backslash j }^n, F_j^{n-1}\right) 
\\& - I \left( Y_j^{n-1}, F_j^{n-1} ; F_{jn} | W_j,F_{\mathcal{A}\backslash j }^n, F_j^{n-1} \right)
\\& +  H \left( Y_{jn} | W_j,F_{\mathcal{A} }^n, Y_j^{n-1} \right)
\\& \stackrel{{\rm{(c)}}}{=}  H \left( Y_j^{n-1}, F_j^{n-1} | W_j,F_{\mathcal{A}\backslash j }^n, F_j^{n-1}\right) 
\\& +  H \left( Y_{jn} | W_j,F_{\mathcal{A} }^n, Y_j^{n-1} \right)
\\& =  H \left( Y_j^{n-1}, F_j^{n-1} | W_j,F_{\mathcal{A}\backslash j }^n, F_j^{n-1}\right) 
\\& +  \left( 1- \delta_{j}\right) H \left( X_{jn} | W_j,F_{\mathcal{A}\backslash j }^n,  F_j^{n-1},  Y_j^{n-1} \right) ,
\end{align*}
where:
(i) the equality in ${\rm{(a)}}$ follows from the chain rule of the entropy;
(ii) the equality in ${\rm{(b)}}$ is due to the definition of mutual information;
(iii) finally, the equality in ${\rm{(c)}}$ is because $F_{jn}$ is independent of the rest of the random variables.
By recursively proceeding in the same way for $H \left( Y_j^{n-1}, F_j^{n-1} | W_j,F_{\mathcal{A}\backslash j }^n, F_j^{n-1}\right) $, we get the result in \eqref{eq:extraLemma}.

By means of similar steps, it is not difficult to prove the result in \eqref{eq:firsteqLemma} and \eqref{eq:extraLemma} for the RY- and X-network. 

We have, $\forall j \in \mathcal{A}= [1:3]$ and with $W_3= W_{[1:2]}$,
	\begin{align*}
	\epsilon & \stackrel{{\rm{(a)}}}{>} I \left( W_j ; Z_j^n, F_{\mathcal{A}}^n \right)
	\\& \stackrel{{\rm{(b)}}}{=}   I \left( W_j ; Z_j^n, F_j^n | F_{{\mathcal{A}} \backslash j}^n \right)
	\\& \stackrel{{\rm{(c)}}}{=}  \sum_{i=1}^n  I \left( W_j ; Z_{ji}, F_{ji} | F_{{\mathcal{A}} \backslash j}^n, F_j^{i-1}, Z_j^{i-1} \right) 
	\\& \stackrel{{\rm{(d)}}}{=}  \sum_{i=1}^n  I \left( W_j ; Z_{ji}| F_{{\mathcal{A}} \backslash j}^n, F_j^{i-1},F_{ji}, Z_j^{i-1} \right) 
%
%
	\\& =  \sum_{i=1}^n \left( 1- \delta_{jE}\right) I \left( W_j ; X_{ji}| F_{{\mathcal{A}} \backslash j}^n, F_j^{i-1}, Z_j^{i-1} \right),
	\end{align*}
where:
(i) the inequality in ${\rm{(a)}}$ is due to the security constraints;
(ii) the equality in ${\rm{(b)}}$ follows from the independence of $W_j$ on $F_{{\mathcal{A}} \backslash j}^n$;
(iii) the equality in ${\rm{(c)}}$ is due to the chain rule of the mutual information;
(iv) finally, the equality in ${\rm{(d)}}$ is due to the independence of $W_j$ on $F_{ji}$.
This proves \eqref{eq:SeceqLemma} for the Y-network.
By means of similar steps, it is not difficult to prove the result in \eqref{eq:SeceqLemma} for the RY- and X-network.

We then have, $\forall j \in \mathcal{A}= [1:3]$, with $W_3= W_{[1:2]} $ and $R_3 = R_1+R_2$,
\begin{align*}
&n R_j  \stackrel{{\rm{(a)}}}{\leq } I \left( W_j ;Y_3^n,  F_{\mathcal{A}}^n \right) + n \epsilon
\\& \stackrel{{\rm{(b)}}}{\leq } I \left( W_j ;Y_j^n,  F_{\mathcal{A}}^n \right) + n \epsilon
\\& \stackrel{{\rm{(c)}}}{ = } I \left( W_j ;Y_j^n,  F_{\mathcal{A}}^n, Z_j^n \right) - I \left( W_j ; Z_j^n | Y_j^n,  F_{\mathcal{A}}^n \right) + n \epsilon
\\& \stackrel{{\rm{(d)}}}{ \leq } I \left( W_j ;Y_j^n,  F_{\mathcal{A}}^n, Z_j^n \right) + n \epsilon
\\& \stackrel{{\rm{(e)}}}{ = } I \left( W_j ;Y_j^n, F_j^n, Z_j^n |  F_{\mathcal{A} \backslash j}^n \right) + n \epsilon
\\& \stackrel{{\rm{(f)}}}{ = } \sum_{i=1}^n I \left( W_j ;Y_{ji}, F_{ji}, Z_{ji} | Y_j^{i-1}, F_j^{i-1}, Z_j^{i-1}, F_{\mathcal{A} \backslash j}^n \right) + n \epsilon
\\& \stackrel{{\rm{(g)}}}{ = } \sum_{i=1}^n I \left( W_j ;Y_{ji}, Z_{ji} | Y_j^{i-1}, F_j^{i-1}, Z_j^{i-1}, F_{\mathcal{A} \backslash j}^n, F_{ji} \right)+ n \epsilon
%
%
\\& = \left( 1\!-\! \delta_{jE}\delta_j \right)\sum_{i=1}^n I \left( W_j ;X_{ji} | Y_j^{i\!-\!1}, F_j^{i\!-\!1}, Z_j^{i\!-\!1}, F_{\mathcal{A} \backslash j}^n \right)+ n \epsilon,
\end{align*}
where: (i) the inequality in ${\rm{(a)}}$ is due to the Fano's inequality;
(ii) the inequality in ${\rm{(b)}}$ follows from the Markov chain $W_j - Y_j^n, F_{\mathcal{A}}^n - Y_3^n$ for $j \in [1:2]$;
(iii) the equality in ${\rm{(c)}}$ is due to the chain rule of the mutual information;
(iv) the inequality in ${\rm{(d)}}$ is because the mutual information is a non-negative quantity;
(v) the equality in ${\rm{(e)}}$ follows from the independence of $W_j$ and $F_{\mathcal{A} \backslash j}^n$;
(vi) the equality in ${\rm{(f)}}$ is due to the chain rule of the mutual information;
(vii) finally, the equality in ${\rm{(g)}}$ follows from the independence of $W_j$ on $F_{ji}$.
This proves \eqref{eq:ThirdeqLemma} for the Y-network.
By means of similar steps, it is not difficult to prove the result in \eqref{eq:ThirdeqLemma} for the RY- and X-network.

Finally, for node $\mathsf{D}$ in the Y-network we have
\begin{align*}
&n (R_1 + R_2)  
\\& \stackrel{{\rm{(a)}}}{\leq } I \left( W_1, W_2 ;Y_3^n,  F_{\mathcal{A}}^n \right) + n \epsilon
\\& \stackrel{{\rm{(b)}}}{ = } I \left( W_1, W_2 ;Y_3^n, F_3^n |  F_{\mathcal{A} \backslash 3}^n \right) + n \epsilon
\\& \stackrel{{\rm{(c)}}}{ = } \sum_{i=1}^n I \left( W_1, W_2 ;Y_{3i}, F_{3i}| Y_3^{i-1}, F_3^{i-1}, F_{\mathcal{A} \backslash 3}^n \right) + n \epsilon
\\& \stackrel{{\rm{(d)}}}{ = } \sum_{i=1}^n I \left( W_1, W_2 ;Y_{3i} | Y_3^{i-1}, F_3^{i-1}, F_{\mathcal{A} \backslash 3}^n, F_{3i} \right) + n \epsilon
%
%
\\& = \left( 1- \delta_3 \right)\sum_{i=1}^n I \left( W_1, W_2 ;X_{3i} | Y_3^{i-1}, F_3^{i-1}, F_{\mathcal{A} \backslash 3}^n \right) + n \epsilon,
\end{align*}
where:
(i) the inequality in ${\rm{(a)}}$ is due to the Fano's inequality;
(ii) the equality in ${\rm{(b)}}$ follows from the independence of the pair $\left(W_1, W_2 \right)$ on $F_{\mathcal{A} \backslash 3}^n$;
(iii) the equality in ${\rm{(c)}}$ is due to the chain rule of the mutual information;
(iv) finally, the equality in ${\rm{(d)}}$ follows from the independence of the pair $\left(W_1, W_2 \right)$ on $F_{3i}$. 
This proves \eqref{eq:FourtheqLemma} for the Y-network.
By means of similar steps, it is not difficult to prove the results in \eqref{eq:FiftheqLemma} and in \eqref{eq:SixtheqLemma} for the RY- and X-network, respectively.

\section{Proof of the converse of Theorem~\ref{eq:CapY}}
\label{app:ConvThY}

For the Y-network, in~\eqref{eq:corrToT} we set $\mathcal{A}=[1:3]$, $\kappa =3$ and $\lambda \in [1:2]$. 
We use~\eqref{eq:corr3} and~\eqref{eq:corr4} for proving the converse of Theorem~\ref{eq:CapY}. 
The RHS of these two expressions is positive as the entropy of a discrete random variable is positive. \\
{\bf{First constraint.}} From \eqref{eq:corr4} we have
\begin{align*}
&n k_\lambda  
\\&= \! \delta_{\lambda\text{E}} \left( 1\!-\! \delta_\lambda \right) \!\sum_{i\!=\!1}^n H \left( X_{\lambda i} \left | \right. Y_\lambda^{i-1}, F_\lambda^{i-1},W_\lambda,F_{\mathcal{A} \backslash \lambda}^n\right)
\\& \geq \! \delta_{\lambda\text{E}} \left( 1\!-\! \delta_\lambda \right) \!\sum_{i\!=\!1}^n H \left( X_{\lambda i} \left | \right. Y_\lambda^{i-1}, Z_\lambda^{i-1}, F_\lambda^{i-1},W_\lambda,F_{\mathcal{A} \backslash \lambda}^n\right)
\\& \stackrel{{\rm{(a)}}}{\geq }\! \delta_{\lambda\text{E}} \left( 1\!-\! \delta_\lambda \right) \!\sum_{i\!=\!1}^n H \left( X_{\lambda i} \left | \right. Y_\lambda^{i\!-\!1}, Z_\lambda^{i\!-\!1}, F_\lambda^{i\!-\!1},W_\lambda,F_{\mathcal{A} \backslash \lambda}^n\right)
\\&- H \left( Y_\lambda^n \left | \right. W_\lambda,F_{\mathcal{A}}^n,Z_\lambda^n \right)
\\& \stackrel{{\rm{(b)}}}{=} \! \left( 1\!-\!\delta_{\lambda\text{E}} \right) \sum_{i\!=\!1}^n I \left(Y_\lambda^{i\!-\!1},F_\lambda^{i\!-\!1}; X_{\lambda i} \left | \right. Z_{\lambda}^{i\!-\!1},F_\lambda^{i\!-\!1},W_\lambda,F_{\mathcal{A} \backslash \lambda}^n \right)
\\& \stackrel{{\rm{(c)}}}{=}\left( 1-\delta_{\lambda\text{E}} \right)
\left [
\sum_{i=1}^n H \left( X_{\lambda i} \left | \right. Z_{\lambda}^{i-1},F_\lambda^{i-1},W_\lambda,F_{\mathcal{A} \backslash \lambda}^n \right)
\right.
\\& \left. - \sum_{i=1}^n H \left( X_{\lambda i}\left | \right. Z_{\lambda}^{i-1},F_\lambda^{i-1},W_\lambda,F_{\mathcal{A} \backslash \lambda}^n,Y_\lambda^{i-1} \right)
\right ]
\\& \stackrel{{\rm{(d)}}}{=}\left( 1-\delta_{\lambda\text{E}} \right)
\left [
\sum_{i=1}^n H \left( X_{\lambda i} \left | \right. Z_{\lambda}^{i-1},F_\lambda^{i-1},F_{\mathcal{A} \backslash \lambda}^n \right) \right.
\\& \left.- 
\sum_{i=1}^n  I \left( X_{\lambda i};W_\lambda | Z_\lambda^{i-1},F_\lambda^{i-1},F_{\mathcal{A} \backslash \lambda}^n\right)
\right.
\\&  \left. -
\sum_{i=1}^n H \left( X_{\lambda i}\left | \right. Z_{\lambda}^{i-1},F_\lambda^{i-1},W_\lambda,F_{\mathcal{A} \backslash \lambda}^n,Y_\lambda^{i-1} \right)
\right ]
\\& \stackrel{{\rm{(e)}}}{\geq}\left( 1-\delta_{\lambda\text{E}} \right)
\left [
\sum_{i=1}^n H \left( X_{\lambda i} \left | \right. Y_\lambda^{i-1},Z_{\lambda}^{i-1},F_\lambda^{i-1},F_{\mathcal{A} \backslash \lambda}^n \right) \right.
\\& \left.
- 
\sum_{i=1}^n  I \left( X_{\lambda i};W_\lambda | Z_\lambda^{i-1},F_\lambda^{i-1},F_{\mathcal{A} \backslash \lambda}^n\right)
\right.
\\&  \left. -
\sum_{i=1}^n H \left( X_{\lambda i}\left | \right. Z_{\lambda}^{i-1},F_\lambda^{i-1},W_\lambda,F_{\mathcal{A} \backslash \lambda}^n,Y_\lambda^{i-1} \right)
\right ]
\\& \stackrel{{\rm{(f)}}}{=}\left( 1-\delta_{\lambda\text{E}} \right)
\left [
\sum_{i=1}^n I \left( X_{\lambda i}; W_\lambda \left | \right. Y_\lambda^{i-1},Z_{\lambda}^{i-1},F_\lambda^{i-1},F_{\mathcal{A} \backslash \lambda}^n \right)
\right.
\\& \left. -  \sum_{i=1}^n  I \left( X_{\lambda i};W_\lambda | Z_\lambda^{i-1},F_\lambda^{i-1},F_{\mathcal{A} \backslash \lambda}^n\right)
\right ]
\\& \stackrel{{\rm{(g)}}}{\geq}  n R_\lambda \frac{1-\delta_{\lambda\text{E}} }{1-\delta_{\lambda\text{E}} \delta_\lambda} - \epsilon,
\end{align*}
where: 
(i) the inequality in ${\rm{(a)}}$ follows since the entropy of a discrete random variable is positive; 
(ii) the equality in ${\rm{(b)}}$ is due to \eqref{eq:firsteqLemma} in Lemma \ref{lemma:lemma1}; 
(iii) the equality in ${\rm{(c)}}$ is due to the definition of mutual information; 
(iv) the equality in ${\rm{(d)}}$ follows from the definition of mutual information; 
(v) the inequality in ${\rm{(e)}}$ follows from the `conditioning reduces the entropy' principle; 
(vi) the equality in ${\rm{(f)}}$ is due to the definition of mutual information; 
(vii) finally, the inequality in ${\rm{(g)}}$  follows by means of \eqref{eq:SeceqLemma} and \eqref{eq:ThirdeqLemma} in Lemma \ref{lemma:lemma1}. 
\\
{\bf{Second constraint.}} From \eqref{eq:corr3} with $W_{[1:2]} = W_3$ , we have
\begin{align*}
&n k_3  
\\& = \! \delta_{3\text{E}} \left( 1\!-\! \delta_3 \right) \!\sum_{i\!=\!1}^n H \left( X_{3 i} \left | \right. Y_3^{i-1}, Z_3^{i-1}, F_3^{i-1},W_3,F_{\mathcal{A} \backslash 3}^n\right)
\\& \stackrel{{\rm{(a)}}}{\geq }\! \delta_{3\text{E}} \left( 1\!-\! \delta_3 \right) \!\sum_{i\!=\!1}^n H \left( X_{3 i} \left | \right. Y_3^{i\!-\!1}, Z_3^{i\!-\!1}, F_3^{i\!-\!1},W_3,F_{\mathcal{A} \backslash 3}^n\right)
\\&- H \left( Y_3^n \left | \right. W_3,F_{\mathcal{A}}^n,Z_3^n \right)
\\& \stackrel{{\rm{(b)}}}{=} \! \left( 1\!-\!\delta_{3\text{E}} \right) \sum_{i\!=\!1}^n I \left(Y_3^{i\!-\!1},F_3^{i\!-\!1}; X_{3 i} \left | \right. Z_{3}^{i\!-\!1},F_3^{i\!-\!1},W_3,F_{\mathcal{A} \backslash 3}^n \right)
\\& \stackrel{{\rm{(c)}}}{=}\left( 1-\delta_{3\text{E}} \right)
\left [
\sum_{i=1}^n H \left( X_{3 i} \left | \right. Z_{3}^{i-1},F_3^{i-1},W_3,F_{\mathcal{A} \backslash 3}^n \right)
\right.
\\& \left. - \sum_{i=1}^n H \left( X_{3 i}\left | \right. Z_{3}^{i-1},F_3^{i-1},W_3,F_{\mathcal{A} \backslash 3}^n,Y_3^{i-1} \right)
\right ]
\\& \stackrel{{\rm{(d)}}}{=}\left( 1-\delta_{3\text{E}} \right)
\left [
\sum_{i=1}^n H \left( X_{3 i} \left | \right. Z_{3}^{i-1},F_3^{i-1},F_{\mathcal{A} \backslash 3}^n \right) \right.
\\& \left.- 
\sum_{i=1}^n  I \left( X_{3 i};W_3 | Z_3^{i-1},F_3^{i-1},F_{\mathcal{A} \backslash 3}^n\right)
\right.
\\&  \left. -
\sum_{i=1}^n H \left( X_{3 i}\left | \right. Z_{3}^{i-1},F_3^{i-1},W_3,F_{\mathcal{A} \backslash 3}^n,Y_3^{i-1} \right)
\right ]
\\& \stackrel{{\rm{(e)}}}{\geq}\left( 1-\delta_{3\text{E}} \right)
\left [
\sum_{i=1}^n H \left( X_{3 i} \left | \right. Y_3^{i-1},Z_{3}^{i-1},F_3^{i-1},F_{\mathcal{A} \backslash 3}^n \right) \right.
\\& \left.
- 
\sum_{i=1}^n  I \left( X_{3 i};W_3 | Z_3^{i-1},F_3^{i-1},F_{\mathcal{A} \backslash 3}^n\right)
\right.
\\&  \left. -
\sum_{i=1}^n H \left( X_{3 i}\left | \right. Z_{3}^{i-1},F_3^{i-1},W_3,F_{\mathcal{A} \backslash 3}^n,Y_3^{i-1} \right)
\right ]
\\& \stackrel{{\rm{(f)}}}{=}\left( 1-\delta_{3\text{E}} \right)
\left [
\sum_{i=1}^n I \left( X_{3 i}; W_3 \left | \right. Y_3^{i-1},Z_{3}^{i-1},F_3^{i-1},F_{\mathcal{A} \backslash 3}^n \right)
\right.
\\& \left. -  \sum_{i=1}^n  I \left( X_{3 i};W_3 | Z_3^{i-1},F_3^{i-1},F_{\mathcal{A} \backslash 3}^n\right)
\right ]
\\& \stackrel{{\rm{(g)}}}{\geq}  n (R_1 + R_2) \frac{1-\delta_{3\text{E}} }{1-\delta_{3\text{E}} \delta_3} - \epsilon,
\end{align*}
where: 
(i) the inequality in ${\rm{(a)}}$ follows since the entropy of a discrete random variable is positive; 
(ii) the equality in ${\rm{(b)}}$ is due to \eqref{eq:firsteqLemma} in Lemma \ref{lemma:lemma1}; 
(iii) the equality in ${\rm{(c)}}$ is due to the definition of mutual information; 
(iv) the equality in ${\rm{(d)}}$ follows from the definition of mutual information; 
(v) the inequality in ${\rm{(e)}}$ follows from the `conditioning reduces the entropy' principle; 
(vi) the equality in ${\rm{(f)}}$ is due to the definition of mutual information; 
(vii) finally, the inequality in ${\rm{(g)}}$  follows by means of \eqref{eq:SeceqLemma} and \eqref{eq:ThirdeqLemma} in Lemma \ref{lemma:lemma1} with $R_3 = R_1 + R_2$. 
\\
{\bf{Third constraint.}} From \eqref{eq:corr4} we get 

\begin{align*}
\frac{nk_\lambda}{\delta_{\lambda\text{E}}}  & = \left( 1- \delta_{\lambda} \right) \sum_{i=1}^n H \left( X_{\lambda i} \left | \right. Y_{\lambda}^{i-1}, F_{\lambda}^{i-1},W_\lambda, F_{\mathcal{A} \backslash \lambda}^n\right)
\\ & {=} H \left( Y_\lambda^n \left | \right. W_\lambda, F_{\mathcal{A}}^n \right),
\end{align*}
where the last equality follows from \eqref{eq:extraLemma} in Lemma \ref{lemma:lemma1}.

Using this and Fano's inequality with $\lambda \in [1:2]$ we have
\begin{align*}
nR_\lambda + \frac{n k_\lambda}{\delta_{\lambda\text{E}}} & \leq I \left( W_\lambda; Y_3^n, F_{\mathcal{A}}^n \right) + H \left( Y_\lambda^n | W_\lambda,F_{\mathcal{A} }^n \right) + n \epsilon
\\ & \stackrel{{\rm{(a)}}}{\leq} I \left( W_\lambda; Y_\lambda^n, F_{\mathcal{A}}^n \right) + H \left( Y_\lambda^n | W_\lambda,F_{\mathcal{A} }^n \right)  + n \epsilon
\\ & \stackrel{{\rm{(b)}}}{=} I \left( W_\lambda; Y_\lambda^n \left | \right. F_{\mathcal{A}}^n \right) + H \left( Y_\lambda^n | W_\lambda,F_{\mathcal{A} }^n \right) + n \epsilon
\\ & \stackrel{{\rm{(c)}}}{=} H \left( Y_\lambda^n \left | \right. F_{\mathcal{A}}^n \right) + n \epsilon
\\ & \stackrel{{\rm{(d)}}}{\leq} n \left( 1-\delta_\lambda \right) + n \epsilon,
\end{align*}
where: 
(i) the inequality in ${\rm{(a)}}$ follows because of the Markov chain $W_\lambda-Y_\lambda^n, F_{\mathcal{A}}^n-Y_3^n$, $\forall \lambda \in [1:2]$; 
(ii) the equality in ${\rm{(b)}}$ follows because of the independence between $F_{\mathcal{A}}^n$ and $W_\lambda$; 
(iii) the equality in ${\rm{(c)}}$ follows from the definition of mutual information; 
(iv) finally, the inequality in ${\rm{(d)}}$ follows since node $\mathsf{M}$ receives at most $n \left( 1-\delta_\lambda \right)$ packets on channel $\lambda \in [1:2]$.
\\
{\bf{Fourth constraint.}} From \eqref{eq:FourtheqLemma} in Lemma \ref{lemma:lemma1} we have
\begin{align*}
&n \left( R_1+R_2\right)
\\& \leq \left( 1-\delta_3 \right) \sum_{i=1}^n I \left( W_{[1:2]};X_{3i} \left | \right. F_{\mathcal{A} \backslash 3}^n,Y_3^{i-1},F_3^{i-1}\right)
\\ & \stackrel{{\rm{(a)}}}{=} \left( 1-\delta_3 \right) \left [
\sum_{i=1}^n H \left( X_{3i} \left | \right. F_{\mathcal{A} \backslash 3}^n,Y_3^{i-1},F_3^{i-1} \right) \right.
\\& \left. -\sum_{i=1}^n H \left( X_{3i} \left | \right. F_{\mathcal{A} \backslash 3}^n,Y_3^{i-1},F_3^{i-1},W_{[1:2]} \right)
\right ]
\\ & \stackrel{{\rm{(b)}}}{\leq} \left( 1-\delta_3 \right) n 
\\&- \left( 1-\delta_3 \right) \sum_{i=1}^n H \left( X_{3i} \left | \right. F_{\mathcal{A} \backslash 3}^n,Y_3^{i-1},F_3^{i-1},W_{[1:2]} \right)
\\ & \stackrel{{\rm{(c)}}}{\leq} \left( 1-\delta_3 \right) n 
\\& - \left( 1-\delta_3 \right) \sum_{i=1}^n H \left( X_{3i} \left | \right. F_{\mathcal{A} \backslash 3}^n,Y_3^{i-1},F_3^{i-1},W_{[1:2]},Z_3^{i-1} \right)
\\ & \stackrel{{\rm{(d)}}}{=} \left( 1-\delta_3 \right) n - \frac{k_3}{\delta_{3\text{E}}} n,
\end{align*}
where: 
(i) the equality in ${\rm{(a)}}$ follows from the definition of mutual information; 
(ii) the inequality in ${\rm{(b)}}$ is due to the fact that $H\left( X_{3i}\right) \leq 1$; 
(iii) the inequality in ${\rm{(c)}}$ is due to the `conditioning reduces the entropy' principle; 
(iv) finally, the equality in ${\rm{(d)}}$ follows from \eqref{eq:corr3}.
\\
{\bf{Fifth constraint.}} From \eqref{eq:extraLemma} and \eqref{eq:corr4} we have
\begin{align*}
& n \left (\frac{k_1}{\delta_{1\text{E}}}+ \frac{k_2}{\delta_{2\text{E}}} \right ) \frac{(1-\delta_3) \delta_{3\text{E}}}{1-\delta_3 \delta_{3\text{E}}} 
\\&=
\left [
H \left( Y_1^n | W_1,F_{\mathcal{A} }^n \right) + H \left( Y_2^n | W_2,F_{\mathcal{A} }^n \right)
\right ]\frac{(1-\delta_3) \delta_{3\text{E}}}{1-\delta_3 \delta_{3\text{E}}}
\\ & \stackrel{{\rm{(a)}}}{\geq} \left [
H \left( Y_1^n | W_{[1,2]},F_{\mathcal{A} }^n \right) + H \left( Y_2^n | W_{[1,2]},F_{\mathcal{A} }^n \right)
\right ]\frac{(1-\delta_3) \delta_{3\text{E}}}{1-\delta_3 \delta_{3\text{E}}}
\\& \stackrel{{\rm{(b)}}}{\geq}  H \left( Y_1^n,Y_2^n | W_{[1,2]},F_{\mathcal{A} }^n \right) \frac{(1-\delta_3) \delta_{3\text{E}}}{1-\delta_3 \delta_{3\text{E}}}
\\& \stackrel{{\rm{(c)}}}{\geq} I \left( Y_1^n,Y_2^n;Y_3^n,Z_3^n | W_{[1,2]},F_{\mathcal{A} }^n \right) \frac{(1-\delta_3) \delta_{3\text{E}}}{1-\delta_3 \delta_{3\text{E}}}
\\ & \stackrel{{\rm{(d)}}}{=} \frac{(1-\delta_3) \delta_{3\text{E}}}{1-\delta_3 \delta_{3\text{E}}} \cdot
\\& \sum_{i=1}^n I \left( Y_1^n, Y_2^n; Y_{3i},Z_{3i} \left | \right.
W_{[1,2]},F_{\mathcal{A}}^n,Y_3^{i-1},Z_3^{i-1}\right)
\\& = (1-\delta_3) \delta_{3\text{E}} \cdot
\\& \sum_{i=1}^n I \left( Y_1^n, Y_2^n; X_{3i} \left | \right.
W_{[1,2]},F_{\mathcal{A}\backslash 3 }^n,Y_3^{i-1},F_3^{i-1},Z_3^{i-1},F_{3i}^n \right)
\\ & \stackrel{{\rm{(e)}}}{=}  (1-\delta_3) \delta_{3\text{E}} \cdot
\\& \sum_{i=1}^n H \left( X_{3i} \left | \right.
W_{[1,2]},F_{\mathcal{A}\backslash 3 }^n,Y_3^{i-1},F_3^{i-1},Z_3^{i-1},F_{3i}^n \right)
\\ & \stackrel{{\rm{(f)}}}{=}  (1\!-\!\delta_3) \delta_{3\text{E}}\sum_{i=1}^n H \left( X_{3i} \left | \right.
W_{[1,2]},F_{\mathcal{A}\backslash 3 }^n,Y_3^{i-1},F_3^{i-1},Z_3^{i-1} \right)
\\ & \stackrel{{\rm{(g)}}}{=} n k_3,
\end{align*}
where: 
(i) the inequality in ${\rm{(a)}}$ is due to the `conditioning reduces the entropy' principle; 
(ii) the inequality in ${\rm{(b)}}$ follows since $H \left( X,Y\right) \leq H(X) + H(Y)$; 
(iii) the inequality in ${\rm{(c)}}$ follows since the entropy of a discrete random variable is a positive quantity; 
(iv)
the equality in ${\rm{(d)}}$ is due to the chain rule of the mutual information;  
(v) the equality in ${\rm{(e)}}$ follows since node $\mathsf{M}$ does not have any randomness and so $X_{3i}$ is uniquely determined by knowing $\left( Y_1^n,Y_2^n,F_{\mathcal{A}}^n\right)$; 
(vi) the equality in ${\rm{(f)}}$ is due to the Markov chain $X_{3i} - W_1,W_2,F_{\mathcal{A}\backslash 3 }^n,Y_3^{i-1},F_3^{i-1},Z_3^{i-1} - F_{3i}^n$; 
(vii) finally, the equality in ${\rm{(g)}}$ follows from \eqref{eq:corr3}.

\section{Proof of the converse of Theorem~\ref{eq:CapInvY}}
\label{app:ConvThRY}
For the RY-network, in~\eqref{eq:corrToT}  we set 
$\mathcal{A}=[1:3]$, $j = 3$, $\kappa \in [1:2]$, $\mathcal{B} =[1:2]$ and $e_3=e$.
We start by proving that the Righ-Hand Side (RHS) of the quantities in~\eqref{eq:corrToT} is positive.
We use~\eqref{eq:corr1},~\eqref{eq:corr2} and~\eqref{eq:corr3} for proving the converse of Theorem~\ref{eq:CapInvY}. 
The RHS of~\eqref{eq:corr3} is positive as the entropy of a discrete random variable is positive. 
For the RHS of \eqref{eq:corr1}, we have
\begin{align*}
& \left( 1- \delta_3 \delta_{3\text{E}} \right) \sum_{i=1}^n H \left( X_{3i} \left | \right. Y_3^{i-1}, Z_3^{i-1}, F_3^{i-1},W_{[1:2]},F_{\mathcal{A} \backslash 3}^n\right) 
\\& - H \left( Y_3^n | W_{[1:2]},F_{\mathcal{A} }^n \right)
\\& \stackrel{{\rm{(a)}}}{=}  \sum_{i=1}^n H \left( Y_{3i}, Z_{3i} \left | \right. Y_3^{i-1}, Z_3^{i-1}, F_3^n,W_{[1:2]},F_{\mathcal{A} \backslash 3}^n\right) 
\\&- H \left( Y_3^n | W_{[1:2]},F_{\mathcal{A} }^n \right)
\\& \stackrel{{\rm{(b)}}}{=}  H \left( Y_3^n, Z_3^n \left | \right. F_3^n,W_{[1:2]},F_{\mathcal{A} \backslash 3}^n\right) - H \left( Y_3^n | W_{[1:2]},F_{\mathcal{A} }^n \right)
\\&= H \left( Y_3^n, Z_3^n \left | \right. W_{[1:2]},F_{\mathcal{A}}^n\right) - H \left( Y_3^n | W_{[1:2]},F_{\mathcal{A} }^n \right)
\geq  0,
\end{align*}
where: 
(i) the equality in ${\rm{(a)}}$ follows because given $F_{3i}$, the pair $\left(Y_{3i}, Z_{3i} \right)$ is equal to $X_{3i}$ with probability $\left( 1- \delta_3 \delta_{3\text{E}} \right)$ and null otherwise and because of the Markov chain $X_{3i} -W_{[1:2]},F_{\mathcal{A} \backslash 3}^n,Y_3^{i-1},Z_3^{i-1},F_3^{i-1} - F_{3i}^n$; 
(ii) finally, the equality in ${\rm{(b)}}$ is due to the chain rule of entropy. 

For the RHS of \eqref{eq:corr2}, we have
\begin{align*}
&H \left( Y_3^n | W_{[1:2]},F_{\mathcal{A} }^n \right) 
\\& - \left( 1-\delta_3\right) \sum_{i=1}^n H \left( X_{3i}\left | \right. Y_3^{i-1},Z_3^{i-1},F_{\mathcal{A}\backslash 3 }^n,F_3^{i-1},W_{[1:2]} \right)
\\& \stackrel{{\rm{(a)}}}{=} \sum_{i=1}^n H \left( Y_{3i} | W_{[1:2]},F_{\mathcal{A} }^n,Y_3^{i-1} \right) 
\\& - \left( 1-\delta_3\right) \sum_{i=1}^n H \left( X_{3i}\left | \right. Y_3^{i-1},Z_3^{i-1},F_{\mathcal{A}\backslash 3}^n,F_3^{i-1},W_{[1:2]} \right)
\\& \stackrel{{\rm{(b)}}}{\geq} \sum_{i=1}^n H \left( Y_{3i} | W_{[1:2]},F_{\mathcal{A} }^n,Y_3^{i-1},Z_3^{i-1} \right) 
\\&- \left( 1-\delta_3\right) \sum_{i=1}^n H \left( X_{3i}\left | \right. Y_3^{i-1},Z_3^{i-1},F_{\mathcal{A}\backslash 3}^n,F_3^{i-1},W_{[1:2]} \right)
\\&=\! \left( 1\!-\!\delta_3\right)  \left [\sum_{i\!=\!1}^n H \left( X_{3i} | W_{[1:2]},F_{\mathcal{A} \backslash 3}^n,Y_3^{i\!-\!1},Z_3^{i-1},F_3^{i\!-\!1},F_{3i}^{n} \right) \right.
\\ &\left.   - \sum_{i=1}^n H \left( X_{3i}\left | \right. Y_3^{i-1},Z_3^{i-1},F_{\mathcal{A}\backslash 3 }^n,F_3^{i-1},W_{[1:2]} \right) \right ]
\\& \stackrel{{\rm{(c)}}}{=} \left( 1-\delta_3\right)  \left [\sum_{i=1}^n H \left( X_{3i} | W_{[1:2]},F_{\mathcal{A} \backslash 3}^n,Y_3^{i-1},Z_3^{i-1},F_3^{i-1} \right) \right.
\\& \left.  - \sum_{i=1}^n H \left( X_{3i}\left | \right. Y_3^{i-1},Z_3^{i-1},F_{\mathcal{A}\backslash 3 }^n,F_3^{i-1},W_{[1:2]} \right) \right ] = 0,
\end{align*}
where: 
(i) the equality in ${\rm{(a)}}$ follows from the chain rule of the entropy; (ii) the inequality in ${\rm{(b)}}$ is due to the conditioning reduces the entropy principle; 
(iii) finally, the equality in ${\rm{(c)}}$ follows because of the Markov chain $X_{3i} -W_{[1:2]},F_{\mathcal{A} \backslash 3}^n,Y_3^{i-1},Z_3^{i-1},F_3^{i-1} - F_{3i}^n$.
\\
{\bf{First constraints.}} From \eqref{eq:corr1} and \eqref{eq:corr2} we have
\begin{align*}
&n k_3 + n e \frac{(1-\delta_3) \delta_{3 \text{E}}}{1-\delta_3 \delta_{3\text{E}}} 
\\&= \delta_{3\text{E}} \left( 1- \delta_3 \right) \sum_{i\!=\!1}^n H \left( X_{3i} \left | \right. Y_3^{i-1}, Z_3^{i\!-\!1}, F_3^{i\!-\!1},W_{[1:2]},F_{\mathcal{A} \backslash 3}^n\right)
\\& \stackrel{{\rm{(a)}}}{\geq }\delta_{3\text{E}} \left( 1\!-\! \delta_3 \right) \sum_{i\!=\!1}^n H \left( X_{3i} \left | \right. Y_3^{i\!-\!1}, Z_3^{i\!-\!1}, F_3^{i\!-\!1},W_{[1:2]},F_{\mathcal{A} \backslash 3}^n\right)
\\& - H \left( Y_3^n \left | \right. W_{[1:2]},F_{\mathcal{A}}^n,Z_3^n \right)
\\& \stackrel{{\rm{(b)}}}{= } \left( 1-\delta_{3\text{E}} \right) \cdot
\\& \sum_{i=1}^n I \left(Y_3^{i-1},F_3^{i-1}; X_{3i} \left | \right. Z_{3}^{i-1},F_3^{i-1},W_{[1:2]},F_{\mathcal{A} \backslash 3}^n \right)
\\& \stackrel{{\rm{(c)}}}{=}\left( 1-\delta_{3\text{E}} \right)
\left [
\sum_{i=1}^n H \left( X_{3i} \left | \right. Z_{3}^{i-1},F_3^{i-1},W_{[1:2]},F_{\mathcal{A} \backslash 3}^n \right)  \right.
\\& \left.  -
\sum_{i=1}^n H \left( X_{3i}\left | \right. Z_{3}^{i-1},F_3^{i-1},W_{[1:2]},F_{\mathcal{A} \backslash 3}^n,Y_3^{i-1} \right)
\right ]
\\& \stackrel{{\rm{(d)}}}{=}\left( 1-\delta_{3\text{E}} \right)
\left [
\sum_{i=1}^n H \left( X_{3i} \left | \right. Z_{3}^{i-1},F_3^{i-1},F_{\mathcal{A} \backslash 3}^n \right)
\right.
\\&  \left. - \sum_{i=1}^n  I \left( X_{3i};W_{[1:2]} | Z_3^{i-1},F_3^{i-1},F_{\mathcal{A} \backslash 3}^n\right)
\right.
\\&  \left. - \sum_{i=1}^n H \left( X_{3i}\left | \right. Z_{3}^{i-1},F_3^{i-1},W_{[1:2]},F_{\mathcal{A} \backslash 3}^n,Y_3^{i-1} \right)
\right ]
\\& \stackrel{{\rm{(e)}}}{\geq}\left( 1-\delta_{3\text{E}} \right)
\left [
\sum_{i=1}^n H \left( X_{3i} \left | \right. Y_3^{i-1},Z_{3}^{i-1},F_3^{i-1},F_{\mathcal{A} \backslash 3}^n \right) 
\right. 
\\&  \left. - \sum_{i=1}^n  I \left( X_{3i};W_{[1:2]} | Z_3^{i-1},F_3^{i-1},F_{\mathcal{A} \backslash 3}^n\right)
\right.
\\&  \left. - \sum_{i=1}^n H \left( X_{3i}\left | \right. Z_{3}^{i-1},F_3^{i-1},W_{[1:2]},F_{\mathcal{A} \backslash 3}^n,Y_3^{i-1} \right)
\right ]
\\& \stackrel{{\rm{(f)}}}{=}\left( 1-\delta_{3\text{E}} \right)
\left [
\sum_{i=1}^n I \left( X_{3i}; W_{[1:2]} \left | \right. Y_3^{i-1},Z_{3}^{i-1},F_3^{i-1},F_{\mathcal{A} \backslash 3}^n \right)  \right.
\\& \left. 
-\sum_{i=1}^n  I \left( X_{3i};W_{[1:2]} | Z_3^{i-1},F_3^{i-1},F_{\mathcal{A} \backslash 3}^n\right)
\right ]
\\& \stackrel{{\rm{(g)}}}{\geq}  n (R_1+R_2) \frac{1-\delta_{3\text{E}} }{1-\delta_{3\text{E}} \delta_3} - \epsilon,
\end{align*}
where: 
(i) the inequality in ${\rm{(a)}}$ follows since the entropy of a discrete random variable is positive; 
(ii) the equality in ${\rm{(b)}}$ is due to \eqref{eq:firsteqLemma} in Lemma \ref{lemma:lemma1}; 
(iii) the equality in ${\rm{(c)}}$ is due to the definition of mutual information; 
(iv) the equality in ${\rm{(d)}}$ follows from the definition of mutual information; 
(v) the inequality in ${\rm{(e)}}$ follows from the `conditioning reduces the entropy' principle; 
(vi) the equality in ${\rm{(f)}}$ is due to the definition of mutual information; 
(vii) finally, the inequality in ${\rm{(g)}}$ follows by means of \eqref{eq:SeceqLemma} and \eqref{eq:ThirdeqLemma} in Lemma \ref{lemma:lemma1}.
\\
{\bf{Second constraints.}} From \eqref{eq:corr3} we have
\begin{align*}
&n k_{\kappa} 
\\&= \delta_{\kappa \text{E}} \left( 1\!-\! \delta_{\kappa} \right) \sum_{i\!=\!1}^n H \left( X_{\kappa i} \left | \right. Y_{\kappa}^{i\!-\!1}, Z_{\kappa}^{i\!-\!1}, F_{\kappa}^{i\!-\!1},W_{[1:2]},F_{\mathcal{A} \backslash \kappa}^n\right)
\\& \stackrel{{\rm{(a)}}}{\geq }\delta_{\kappa \text{E}} \left( 1\!-\! \delta_{\kappa} \right) \sum_{i\!=\!1}^n H \left( X_{\kappa i} \left | \right. Y_{\kappa}^{i\!-\!1}, Z_{\kappa}^{i\!-\!1}, F_{\kappa}^{i\!-\!1},W_{[1:2]},F_{\mathcal{A} \backslash \kappa}^n\right)
\\& - H \left( Y_{\kappa}^n \left | \right. W_{[1:2]},F_{\mathcal{A}}^n,Z_{\kappa}^n \right)
\\& \stackrel{{\rm{(b)}}}{= } \left( 1-\delta_{\kappa \text{E}} \right) \\& \sum_{i=1}^n I \left(Y_{\kappa}^{i-1},F_{\kappa}^{i-1}; X_{\kappa i} \left | \right. Z_{\kappa}^{i-1},F_{\kappa}^{i-1},W_{[1:2]},F_{\mathcal{A} \backslash \kappa}^n \right)
\\& \stackrel{{\rm{(c)}}}{=}\left( 1-\delta_{\kappa \text{E}} \right)
\left [
\sum_{i=1}^n H \left( X_{\kappa i} \left | \right. Z_{\kappa}^{i-1},F_{\kappa}^{i-1},W_{[1:2]},F_{\mathcal{A} \backslash \kappa}^n \right)  \right.
\\& \left. -
\sum_{i=1}^n H \left( X_{\kappa i}\left | \right. Z_{\kappa}^{i-1},F_{\kappa}^{i-1},W_{[1:2]},F_{\mathcal{A} \backslash \kappa}^n,Y_{\kappa}^{i-1} \right)
\right ]
\\& \stackrel{{\rm{(d)}}}{=}\left( 1-\delta_{\kappa \text{E}} \right)
\left [
\sum_{i=1}^n H \left( X_{\kappa i} \left | \right. Z_{\kappa}^{i-1},F_{\kappa}^{i-1},F_{\mathcal{A} \backslash \kappa}^n \right) \right.
\\ & \left. - 
\sum_{i=1}^n  I \left( X_{\kappa i};W_{[1:2]} | Z_{\kappa}^{i-1},F_{\kappa}^{i-1},F_{\mathcal{A} \backslash \kappa}^n\right)
\right.
\\&  \left. -
\sum_{i=1}^n H \left( X_{\kappa i}\left | \right. Z_{\kappa}^{i-1},F_{\kappa}^{i-1},W_{[1:2]},F_{\mathcal{A} \backslash \kappa}^n,Y_{\kappa}^{i-1} \right)
\right ]
\\& \stackrel{{\rm{(e)}}}{\geq}\left( 1-\delta_{\kappa \text{E}} \right)
\left [
\sum_{i=1}^n H \left( X_{\kappa i} \left | \right. Y_{\kappa}^{i-1},Z_{\kappa}^{i-1},F_{\kappa}^{i-1},F_{\mathcal{A} \backslash \kappa}^n \right) \right.
\\& \left. - 
\sum_{i=1}^n  I \left( X_{\kappa i};W_{[1:2]} | Z_{\kappa}^{i-1},F_{\kappa}^{i-1},F_{\mathcal{A} \backslash \kappa}^n\right)
\right.
\\& \left. -
\sum_{i=1}^n H \left( X_{\kappa i}\left | \right. Z_{\kappa}^{i-1},F_{\kappa}^{i-1},W_{[1:2]},F_{\mathcal{A} \backslash \kappa}^n,Y_{\kappa}^{i-1} \right)
\right ]
\\& \stackrel{{\rm{(f)}}}{=}\left( 1-\delta_{\kappa \text{E}} \right)
\left [
\sum_{i=1}^n I \left( X_{\kappa i}; W_{[1:2]} \left | \right. Y_{\kappa}^{i-1},Z_{\kappa}^{i-1},F_{\kappa}^{i-1},F_{\mathcal{A} \backslash \kappa}^n \right)  \right.
\\& \left. 
-\sum_{i=1}^n  I \left( X_{\kappa i};W_{[1:2]} | Z_{\kappa}^{i-1},F_{\kappa}^{i-1},F_{\mathcal{A} \backslash \kappa}^n\right)
\right ]
\\& \stackrel{{\rm{(g)}}}{\geq}  n R_{\kappa} \frac{1-\delta_{\kappa \text{E}} }{1-\delta_{\kappa \text{E}} \delta_{\kappa}} - \epsilon,
\end{align*}
where: 
(i) the inequality in ${\rm{(a)}}$ follows since the entropy of a discrete random variable is positive; 
(ii) the equality in ${\rm{(b)}}$ is due to \eqref{eq:firsteqLemma} in Lemma \ref{lemma:lemma1}; 
(iii) the equality in ${\rm{(c)}}$ is due to the definition of mutual information; 
(iv) the equality in ${\rm{(d)}}$ follows from the definition of mutual information; 
(v) the inequality in ${\rm{(e)}}$ follows from the `conditioning reduces the entropy' principle; 
(vi) the equality in ${\rm{(f)}}$ is due to the definition of mutual information; 
(vii) finally, the inequality in ${\rm{(g)}}$ follows by means by means of \eqref{eq:SeceqLemma} and \eqref{eq:ThirdeqLemma} in Lemma~\ref{lemma:lemma1}.
\\
{\bf{Third constraint.}} From \eqref{eq:corr1} and \eqref{eq:corr2} we get 
$n e +  \frac{n k_3}{\delta_{3\text{E}}} $ = $ H \left( Y_3^n | W_{[1:2]},F_{\mathcal{A} }^n \right) $. 
Now, with this and by using Fano's inequality (keeping in mind that the messages are independent) we have 
\begin{align*}
&n \left( R_1 +R_2\right ) + n e +  \frac{n k_3}{\delta_{3\text{E}}} 
\\& \leq I \left( W_{[1:2]}; Y_1^n,Y_2^n, F_{\mathcal{A}}^n \right) + H \left( Y_3^n | W_{[1:2]},F_{\mathcal{A} }^n \right) + n \epsilon
\\ & \stackrel{{\rm{(a)}}}{\leq} I \left( W_{[1:2]}; Y_3^n, F_{\mathcal{A}}^n \right) + H \left( Y_3^n | W_{[1:2]},F_{\mathcal{A} }^n \right) + n \epsilon
\\ & \stackrel{{\rm{(b)}}}{=} I \left( W_{[1:2]}; Y_3^n \left | \right. F_{\mathcal{A}}^n \right) + H \left( Y_3^n | W_{[1:2]},F_{\mathcal{A} }^n \right) + n \epsilon
\\ & \stackrel{{\rm{(c)}}}{=} H \left( Y_3^n \left | \right. F_{\mathcal{A}}^n \right)+ n \epsilon
\\ & \stackrel{{\rm{(d)}}}{\leq} n \left( 1-\delta_3 \right)+ n \epsilon,
\end{align*}
where: 
(i) the inequality in ${\rm{(a)}}$ follows because of the Markov chain $W_{[1:2]}-Y_3^n, F_{\mathcal{A}}^n-Y_1^n,Y_2^n$; 
(ii) the equality in ${\rm{(b)}}$ follows because of the independence between $F_{\mathcal{A}}^n$ and $\left(W_1,W_2 \right)$; 
(iii) the equality in ${\rm{(c)}}$ follows from the definition of mutual information; 
(iv) finally, the inequality in ${\rm{(d)}}$ follows since node $\mathsf{M}$ receives at most $n \left( 1-\delta_3 \right)$ packets.
\\
{\bf{Fourth constraint.}} From \eqref{eq:FiftheqLemma} in Lemma \ref{lemma:lemma1} we have
\begin{align*}
n  R_{\kappa} &\leq \left( 1-\delta_{\kappa} \right) \sum_{i=1}^n I \left( W_{\kappa};X_{{\kappa}i} \left | \right. F_{\mathcal{A} \backslash {\kappa}}^n,Y_{\kappa}^{i-1},F_{\kappa}^{i-1}\right)
\\ & \stackrel{{\rm{(a)}}}{=} \left( 1-\delta_{\kappa} \right) \left [
\sum_{i=1}^n H \left( X_{{\kappa}i} \left | \right. F_{\mathcal{A} \backslash {\kappa}}^n,Y_{\kappa}^{i-1},F_{\kappa}^{i-1} \right) \right.
\\& \left.
-\sum_{i=1}^n H \left( X_{{\kappa}i} \left | \right. F_{\mathcal{A} \backslash {\kappa}}^n,Y_{\kappa}^{i-1},F_{\kappa}^{i-1},W_{\kappa} \right)
\right ]
\\ & \stackrel{{\rm{(b)}}}{\leq} \left( 1-\delta_{\kappa} \right) n 
\\& - \left( 1-\delta_{\kappa} \right) \sum_{i=1}^n H \left( X_{{\kappa}i} \left | \right. F_{\mathcal{A} \backslash {\kappa}}^n,Y_{\kappa}^{i-1},F_{\kappa}^{i-1},W_{\kappa} \right)
\\ & \stackrel{{\rm{(c)}}}{\leq} \left( 1-\delta_{\kappa} \right) n 
\\&- \left( 1-\delta_{\kappa} \right) \sum_{i=1}^n H \left( X_{{\kappa}i} \left | \right. F_{\mathcal{A} \backslash {\kappa}}^n,Y_{\kappa}^{i-1},F_{\kappa}^{i-1},W_{[1:2]},Z_{\kappa}^{i-1} \right)
\\ & \stackrel{{\rm{(d)}}}{=} \left( 1-\delta_{\kappa} \right) n - \frac{k_{\kappa}}{\delta_{{\kappa}\text{E}}} n,
\end{align*}
where: 
(i) the equality in ${\rm{(a)}}$ follows from the definition of mutual information; 
(ii) the inequality in ${\rm{(b)}}$ is due to the fact that $H\left( X_{\kappa i}\right) \leq 1$; 
(iii) the inequality in ${\rm{(c)}}$ is due to the `conditioning reduces the entropy' principle; 
(iv) finally, the equality in ${\rm{(d)}}$ follows from \eqref{eq:corr3}.
\\
{\bf{Fifth constraint.}} The node $\mathsf{S}$ has a discrete source of randomness $U_0$ such that
\begin{align*}
& n D_0 = H \left( U_0\right)
\\ & \stackrel{{\rm{(a)}}}{=} H \left( U_0 \left| \right. W_{[1:2]}, F_{\mathcal{A}\backslash 3}^n\right)
\\ & \stackrel{{\rm{(b)}}}{\geq} I \left( U_0;Y_3^n,Z_3^n,F_3^n \left| \right. W_{[1:2]}, F_{\mathcal{A}\backslash 3}^n\right)
\\ & \stackrel{{\rm{(c)}}}{=} \!\sum_{i=1}^n I \left( U_0;Y_{3i},Z_{3i},F_{3i} \left| \right. W_{[1:2]}, F_{\mathcal{A}\backslash 3}^n,Y_3^{i\!-\!1},Z_3^{i\!-\!1},F_3^{i\!-\!1}\right)
\\ & \stackrel{{\rm{(d)}}}{=} \! \sum_{i=1}^n I \left( U_0;Y_{3i},Z_{3i} \left| \right.  W_{[1:2]}, F_{\mathcal{A}\backslash 3}^n,Y_3^{i\!-\!1},Z_3^{i\!-\!1},F_3^{i\!-\!1},F_{3i}\right)
\\& = \left( 1-\delta_{3\text{E}} \delta_3 \right) \cdot
\\& \sum_{i=1}^n I \left( U_0;X_{3i} \left| \right.  W_{[1:2]}, F_{\mathcal{A}\backslash 3}^n,Y_3^{i-1},Z_3^{i-1},F_3^{i-1}\right)
\\& \stackrel{{\rm{(e)}}}{=} \left( 1\!-\!\delta_{3\text{E}} \delta_3 \right) \sum_{i\!=\!1}^n H \left( X_{3i} \left| \right.  W_{[1:2]}, F_{\mathcal{A}\backslash 3}^n,Y_3^{i\!-\!1},Z_3^{i\!-\!1},F_3^{i\!-\!1}\right)
%
\\& \stackrel{{\rm{(f)}}}{=} \frac{1-\delta_{3\text{E}} \delta_3}{\delta_{3\text{E}} \left( 1-\delta_3\right)} \left( nk_3 + n e \frac{(1-\delta_3) \delta_{3 \text{E}}}{1-\delta_3 \delta_{3\text{E}}} \right)
\\& = \frac{1-\delta_{3\text{E}} \delta_3}{\delta_{3\text{E}} \left( 1-\delta_3\right)} n k_3 + n e,
\end{align*}
where: 
(i) the equality in ${\rm{(a)}}$ is due to independence of $U_0$ on the rest of the random variables; 
(ii) the inequality in ${\rm{(b)}}$ follows since the entropy of a discrete random variable is positive; 
(iii) the equality in ${\rm{(c)}}$ is due to the chain rule of the mutual information; 
(iv) the equality in ${\rm{(d)}}$ follows because of the independence between $F_{{3i}}$ and $U_0$; 
(v) the equality in ${\rm{(e)}}$ follows because $X_{3i}$ is uniquely determined given $\left( U_0,W_1,W_2,F_{\mathcal{A}}^{i-1}\right)$; 
(vi) finally, the equality in ${\rm{(f)}}$ follows from \eqref{eq:corr1} and \eqref{eq:corr2}.
\\
{\bf{Sixth constraint.}} From \eqref{eq:corr1} and \eqref{eq:corr2} we have
\begin{align*}
& n \left (e + \frac{k_3}{\delta_{3 \text{E}}} \right ) \frac{(1-\delta_{\kappa}) \delta_{{\kappa} \text{E}}}{1-\delta_{\kappa} \delta_{\kappa \text{E}}}
\\& = \frac{(1-\delta_{\kappa}) \delta_{\kappa \text{E}}}{1-\delta_{\kappa} \delta_{\kappa \text{E}}} H \left( Y_3^n | W_{[1:2]},F_{\mathcal{A} }^n \right)
\\& \stackrel{{\rm{(a)}}}{\geq} \frac{(1-\delta_{\kappa}) \delta_{\kappa \text{E}}}{1-\delta_{\kappa} \delta_{\kappa \text{E}}} I \left( Y_3^n; Z_{\kappa}^n, Y_{\kappa}^n| W_{[1:2]},F_{\mathcal{A} }^n \right)
\\& \stackrel{{\rm{(b)}}}{=}  \frac{(1-\delta_{\kappa}) \delta_{\kappa \text{E}}}{1-\delta_{\kappa} \delta_{\kappa \text{E}}} \sum_{i=1}^n I \left( Y_3^n; Z_{\kappa i}, Y_{\kappa i}| W_{[1:2]},F_{\mathcal{A} }^n, Z_{\kappa}^{i-1},Y_{\kappa}^{i-1} \right)
\\& = \frac{(1-\delta_{\kappa}) \delta_{\kappa \text{E}}}{1-\delta_{\kappa} \delta_{\kappa \text{E}}} \left( 1-\delta_{\kappa \text{E}} \delta_{\kappa} \right) \cdot
\\& \sum_{i=1}^n I \left( Y_3^n; X_{\kappa i}| W_{[1:2]},F_{\mathcal{A}\backslash \kappa }^n, Z_{\kappa}^{i-1},Y_{\kappa}^{i-1},F_{\kappa}^{i-1},F_{\kappa i}^n \right)
\\& \stackrel{{\rm{(c)}}}{=}  \frac{(1-\delta_{\kappa}) \delta_{\kappa \text{E}}}{1-\delta_{\kappa} \delta_{\kappa \text{E}}} \left( 1-\delta_{\kappa \text{E}} \delta_{\kappa} \right) \cdot
\\& \sum_{i=1}^n H \left( X_{\kappa i}| W_{[1:2]},F_{\mathcal{A}\backslash \kappa}^n, Z_{\kappa}^{i-1},Y_{\kappa}^{i-1},F_{\kappa}^{i-1},F_{\kappa i}^n \right)
\\& \stackrel{{\rm{(d)}}}{=}  \frac{(1-\delta_{\kappa}) \delta_{\kappa \text{E}}}{1-\delta_{\kappa} \delta_{\kappa \text{E}}} \left( 1-\delta_{\kappa \text{E}} \delta_{\kappa} \right) \cdot
\\& \sum_{i=1}^n H \left( X_{\kappa i}| W_{[1:2]},F_{\mathcal{A}\backslash \kappa}^n, Z_{\kappa}^{i-1},Y_{\kappa}^{i-1},F_{\kappa}^{i-1} \right)
\\& \stackrel{{\rm{(e)}}}{=} n k_{\kappa},
\end{align*}
where: 
(i) the inequality in ${\rm{(a)}}$ follows since the entropy of a discrete random variable is positive; 
(ii) the equality in ${\rm{(b)}}$ is due to the chain rule of the mutual information; 
(iii) the equality in ${\rm{(c)}}$ follows since $X_{\kappa i}$ (with $\kappa \in [1:2]$) is uniquely determined given $\left( Y_3^n, F_{\mathcal{A}}^n\right)$; 
(iv) the equality in ${\rm{(d)}}$ follows because of the Markov chain $X_{\kappa i} -W_{[1:2]},F_{\mathcal{A} \backslash \kappa}^n,Y_{\kappa}^{i-1},Z_{\kappa}^{i-1},F_{\kappa}^{i-1} - F_{\kappa i}^n$; 
(v) finally, the equality in ${\rm{(e)}}$ follows from \eqref{eq:corr3}.

\section{Proof of the converse of Theorem~\ref{eq:CapButterfly}}
\label{app:ConvThX}
For the X-network, in~\eqref{eq:corrToT}  we set $\mathcal{A}=[1:5]$, $\lambda \in [1:2]$, $j = 3$, $\kappa \in [4:5]$, $\mathcal{B} = \{j\}$, $W_3=W_4=W_5 =  W_{[1:2]}$ and $e_3=e$.
We start by proving that the Righ-Hand Side (RHS) of the quantities in~\eqref{eq:corrToT} is positive.
It is straightforward to see that the RHS of \eqref{eq:corr3} and \eqref{eq:corr4} is positive as the entropy of a discrete random variable is positive. 
For the RHS of \eqref{eq:corr1} we have
\begin{align*}
& \left( 1- \delta_j \delta_{j\text{E}} \right) \sum_{i=1}^n H \left( X_{ji} \left | \right. Y_j^{i-1}, Z_j^{i-1}, F_j^{i-1},W_j,F_{\mathcal{A} \backslash j}^n\right) 
\\& - H \left( Y_j^n | W_j,F_{\mathcal{A} }^n \right)
\\& \stackrel{{\rm{(a)}}}{=}  \sum_{i=1}^n H \left( Y_{ji} , Z_{ji} \left | \right. Y_j^{i-1}, Z_j^{i-1}, F_j^n,W_j,F_{\mathcal{A} \backslash j}^n\right) 
\\& - H \left( Y_j^n | W_j,F_{\mathcal{A} }^n \right)
\\& \stackrel{{\rm{(b)}}}{=}  H \left( Y_j^n, Z_j^n \left | \right. F_j^n,W_j,F_{\mathcal{A} \backslash j}^n\right) - H \left( Y_j^n | W_j,F_{\mathcal{A} }^n \right)
\\&= H \left( Y_j^n, Z_j^n \left | \right. W_j,F_{\mathcal{A}}^n\right) - H \left( Y_j^n | W_j,F_{\mathcal{A} }^n \right)
\geq 0,
\end{align*}
where: 
(i) the equality in ${\rm{(a)}}$ follows because given $F_{ji}$, $\left(Y_{ji}, Z_{ji} \right)$ is equal to $X_{ji}$ with probability $\left( 1- \delta_j \delta_{j\text{E}} \right)$ and null otherwise and because of the Markov chain $X_{ji} -W_j,F_{\mathcal{A} \backslash j}^n,Y_j^{i-1},Z_j^{i-1},F_j^{i-1} - F_{ji}^n$; 
(ii) finally, the equality in ${\rm{(b)}}$ is due to the chain rule of entropy. 

For the RHS of \eqref{eq:corr2} we have
\begin{align*}
&H \left( Y_j^n | W_j,F_{\mathcal{A} }^n \right) 
\\&- \left( 1-\delta_j\right) \sum_{i=1}^n H \left( X_{ji}\left | \right. Y_j^{i-1},Z_j^{i-1},F_{\mathcal{A}\backslash j }^n,F_j^{i-1},W_j \right)
\\& \stackrel{{\rm{(a)}}}{=} \sum_{i=1}^n H \left( Y_{ji} | W_j,F_{\mathcal{A} }^n,Y_j^{i-1} \right) 
\\&- \left( 1-\delta_j\right) \sum_{i=1}^n H \left( X_{ji}\left | \right. Y_j^{i-1},Z_j^{i-1},F_{\mathcal{A}\backslash j }^n,F_j^{i-1},W_j \right)
\\& \stackrel{{\rm{(b)}}}{\geq} \sum_{i=1}^n H \left( Y_{ji} | W_j,F_{\mathcal{A} }^n,Y_j^{i-1},Z_j^{i-1} \right) 
\\&- \left( 1-\delta_j\right) \sum_{i=1}^n H \left( X_{ji}\left | \right. Y_j^{i-1},Z_j^{i-1},F_{\mathcal{A}\backslash j }^n,F_j^{i-1},W_j \right)
\\&= \left( 1-\delta_j\right)  \left [\sum_{i=1}^n H \left( X_{ji} | W_j,F_{\mathcal{A} \backslash j}^n,Y_j^{i-1},Z_j^{i-1},F_j^{i-1},F_{ji}^{n} \right)  \right.
\\& \left. \qquad- \sum_{i=1}^n H \left( X_{ji}\left | \right. Y_j^{i-1},Z_j^{i-1},F_{\mathcal{A}\backslash j }^n,F_j^{i-1},W_j \right)\right ]
\\& \stackrel{{\rm{(c)}}}{=} \left( 1-\delta_j\right)  \left [\sum_{i=1}^n H \left( X_{ji} | W_j,F_{\mathcal{A} \backslash j}^n,Y_j^{i-1},Z_j^{i-1},F_j^{i-1} \right)  \right.
\\& \left. \qquad- \sum_{i=1}^n H \left( X_{ji}\left | \right. Y_j^{i-1},Z_j^{i-1},F_{\mathcal{A}\backslash j }^n,F_j^{i-1},W_j \right)\right ] = 0,
\end{align*}
where: 
(i) the equality in ${\rm{(a)}}$ follows from the chain rule of the entropy;
(ii) the inequality in ${\rm{(b)}}$ is due to the `conditioning reduces the entropy' principle; 
(iii) finally, the equality in ${\rm{(c)}}$ follows because of the Markov chain $X_{ji} -W_j,F_{\mathcal{A} \backslash j}^n,Y_j^{i-1},Z_j^{i-1},F_j^{i-1} - F_{ji}^n$.
\\
{\bf{First to third constraints.}} 
From \eqref{eq:corr4}, $\forall j \in [1:2]$
\begin{align*}
&n k_j
\\&= \delta_{j\text{E}} \left( 1- \delta_j \right) \sum_{i=1}^n H \left( X_{ji} \left | \right. Y_j^{i-1}, F_j^{i-1},W_j,F_{\mathcal{A} \backslash j}^n\right)
\\& \geq \delta_{j\text{E}} \left( 1- \delta_j \right) \sum_{i=1}^n H \left( X_{ji} \left | \right. Y_j^{i-1}, Z_j^{i-1}, F_j^{i-1},W_j,F_{\mathcal{A} \backslash j}^n\right),
\end{align*}
where the inequality follows from the `conditioning reduces the entropy' principle.
From \eqref{eq:corr3}, $\forall j \in [4:5]$
\begin{align*}
&n k_j 
\\&= \delta_{j\text{E}} \left( 1- \delta_j \right) \sum_{i=1}^n H \left( X_{ji} \left | \right. Y_j^{i-1}, Z_j^{i-1}, F_j^{i-1},W_j,F_{\mathcal{A} \backslash j}^n\right).
\end{align*}
From \eqref{eq:corr1} and \eqref{eq:corr2} with $j = 3 $
\begin{align*}
&n k_j + n e_j \frac{(1-\delta_j) \delta_{j \text{E}}}{1-\delta_j \delta_{j\text{E}}} 
\\&= \delta_{j\text{E}} \left( 1- \delta_j \right) \sum_{i=1}^n H \left( X_{ji} \left | \right. Y_j^{i-1}, Z_j^{i-1}, F_j^{i-1},W_j,F_{\mathcal{A} \backslash j}^n\right).
\end{align*}
Thus, $\forall j \in [1:5]$ with $R_3 = R_1+R_2$, $R_4 = R_1$, $R_5 = R_2$ and $e_1 = e_ 2 = e_4 = e_5 = 0$,  we have
\begin{align*}
&n k_j + n e_j \frac{(1-\delta_j) \delta_{j \text{E}}}{1-\delta_j \delta_{j\text{E}}} 
\\& \geq \delta_{j\text{E}} \left( 1- \delta_j \right) \sum_{i=1}^n H \left( X_{ji} \left | \right. Y_j^{i-1}, Z_j^{i-1}, F_j^{i-1},W_j,F_{\mathcal{A} \backslash j}^n\right)
\\& \stackrel{{\rm{(a)}}}{\geq }\delta_{j\text{E}} \left( 1- \delta_j \right) \sum_{i=1}^n H \left( X_{ji} \left | \right. Y_j^{i-1}, Z_j^{i-1}, F_j^{i-1},W_j,F_{\mathcal{A} \backslash j}^n\right)
\\&- H \left( Y_j^n \left | \right. W_j,F_{\mathcal{A}}^n,Z_j^n \right)
\\& \stackrel{{\rm{(b)}}}{= } \left( 1\!-\!\delta_{j\text{E}} \right) \sum_{i\!=\!1}^n I \left(Y_j^{i\!-\!1},F_j^{i\!-\!1}; X_{ji} \left | \right. Z_{j}^{i\!-\!1},F_j^{i\!-\!1},W_j,F_{\mathcal{A} \backslash j}^n \right)
\\& \stackrel{{\rm{(c)}}}{=}\left( 1-\delta_{j\text{E}} \right)
\left [
\sum_{i=1}^n H \left( X_{ji} \left | \right. Z_{j}^{i-1},F_j^{i-1},W_j,F_{\mathcal{A} \backslash j}^n \right)
\right.
\\& \left. - \sum_{i=1}^n H \left( X_{ji}\left | \right. Z_{j}^{i-1},F_j^{i-1},W_j,F_{\mathcal{A} \backslash j}^n,Y_j^{i-1} \right)
\right ]
\\& \stackrel{{\rm{(d)}}}{=}\left( 1-\delta_{j\text{E}} \right)
\left [
\sum_{i=1}^n H \left( X_{ji} \left | \right. Z_{j}^{i-1},F_j^{i-1},F_{\mathcal{A} \backslash j}^n \right) \right.
\\& \left. - 
\sum_{i=1}^n  I \left( X_{ji};W_j | Z_j^{i-1},F_j^{i-1},F_{\mathcal{A} \backslash j}^n\right)
\right.
\\&  \left. -
\sum_{i=1}^n H \left( X_{ji}\left | \right. Z_{j}^{i-1},F_j^{i-1},W_j,F_{\mathcal{A} \backslash j}^n,Y_j^{i-1} \right)
\right ]
\\& \stackrel{{\rm{(e)}}}{\geq}\left( 1-\delta_{j\text{E}} \right)
\left [
\sum_{i=1}^n H \left( X_{ji} \left | \right. Y_j^{i-1},Z_{j}^{i-1},F_j^{i-1},F_{\mathcal{A} \backslash j}^n \right) \right.
\\& \left.- 
\sum_{i=1}^n  I \left( X_{ji};W_j | Z_j^{i-1},F_j^{i-1},F_{\mathcal{A} \backslash j}^n\right)
\right.
\\& \left. -
\sum_{i=1}^n H \left( X_{ji}\left | \right. Z_{j}^{i-1},F_j^{i-1},W_j,F_{\mathcal{A} \backslash j}^n,Y_j^{i-1} \right)
\right ]
\\& \stackrel{{\rm{(f)}}}{=}\left( 1-\delta_{j\text{E}} \right)
\left [
\sum_{i=1}^n I \left( X_{ji}; W_j \left | \right. Y_j^{i-1},Z_{j}^{i-1},F_j^{i-1},F_{\mathcal{A} \backslash j}^n \right) \right.
\\& \left. - 
\sum_{i=1}^n  I \left( X_{ji};W_j | Z_j^{i-1},F_j^{i-1},F_{\mathcal{A} \backslash j}^n\right)
\right ]
\\& \stackrel{{\rm{(g)}}}{\geq}  n R_j \frac{1-\delta_{j\text{E}} }{1-\delta_{j\text{E}} \delta_j} - \epsilon,
\end{align*}
where: 
(i) the inequality in ${\rm{(a)}}$ follows since the entropy of a discrete random variable is positive; 
(ii) the equality in ${\rm{(b)}}$ is due  to \eqref{eq:firsteqLemma} in Lemma \ref{lemma:lemma1};
(iii) the equality in ${\rm{(c)}}$ is due to the definition of mutual information; 
(iv) the equality in ${\rm{(d)}}$ follows from the definition of mutual information; 
(v) the inequality in ${\rm{(e)}}$ follows from the `conditioning reduces the entropy' principle; 
(vi) the equality in ${\rm{(f)}}$ is due to the definition of mutual information; 
(vii) finally, the inequality in ${\rm{(g)}}$ follows by means of \eqref{eq:SeceqLemma} and \eqref{eq:ThirdeqLemma} in Lemma \ref{lemma:lemma1}.
\\
{\bf{Fourth constraint.}}
From \eqref{eq:corr4} and \eqref{eq:extraLemma}, we have $\frac{n k_\lambda}{\delta_{\lambda \text{E}}} = H \left( Y_{\lambda}^n | W_{\lambda},F_{\mathcal{A} }^n \right) $. 
By using this and Fano's inequality we have
\begin{align*}
n R_\lambda  +  \frac{n k_\lambda}{\delta_{\lambda\text{E}}}  & \leq I \left( W_\lambda; Y_{\lambda+3}^n, F_{\mathcal{A}}^n \right) + H \left( Y_\lambda^n | W_\lambda,F_{\mathcal{A} }^n \right) 
\\ & \stackrel{{\rm{(a)}}}{\leq} I \left( W_\lambda; Y_3^n, F_{\mathcal{A}}^n \right) + H \left( Y_\lambda^n | W_\lambda,F_{\mathcal{A} }^n \right) + n \epsilon
\\ & \stackrel{{\rm{(b)}}}{\leq} I \left( W_\lambda; Y_\lambda^n, F_{\mathcal{A}}^n \right) + H \left( Y_\lambda^n | W_\lambda,F_{\mathcal{A} }^n \right) + n \epsilon
\\ & \stackrel{{\rm{(c)}}}{=} I \left( W_\lambda; Y_\lambda^n \left | \right. F_{\mathcal{A}}^n \right) + H \left( Y_\lambda^n | W_\lambda,F_{\mathcal{A} }^n \right) + n \epsilon
\\ & \stackrel{{\rm{(d)}}}{=} H \left( Y_\lambda^n \left | \right. F_{\mathcal{A}}^n \right)+ n \epsilon
\\ & \stackrel{{\rm{(e)}}}{\leq} n \left( 1-\delta_\lambda \right)+ n \epsilon,
\end{align*}
where: (i) the inequality in ${\rm{(a)}}$ follows because of the Markov chain $W_\lambda-Y_3^n, F_{\mathcal{A}}^n-Y_{\lambda+3}^n$; (ii) the inequality in ${\rm{(b)}}$ follows because of the Markov chain $W_\lambda-Y_\lambda^n, F_{\mathcal{A}}^n-Y_{3}^n$;
(iii) the equality in ${\rm{(c)}}$ follows because of the independence between $F_{\mathcal{A}}^n$ and $W_\lambda$; (iv) the equality in ${\rm{(d)}}$ follows from the definition of mutual information; (v) finally, the inequality in ${\rm{(e)}}$ follows since node $\mathsf{M}_1$ receives at most $n \left( 1-\delta_\lambda \right)$ packets on channel $\lambda \in [1:2]$.
\\
{\bf{Fifth constraint.}} By means of \eqref{eq:SixtheqLemma} in Lemma \ref{lemma:lemma1} we obtain
\begin{align*}
&n  R_{\lambda}
\\&\leq \left( 1-\delta_{{\lambda}+3} \right) \sum_{i=1}^n I \left( W_{\lambda};X_{{\lambda}+3i} \left | \right. F_{\mathcal{A} \backslash {\lambda}+3}^n,Y_{{\lambda}+3}^{i-1},F_{{\lambda}+3}^{i-1}\right)
\\ & \stackrel{{\rm{(a)}}}{=} \left( 1-\delta_{{\lambda}+3} \right) \left [\sum_{i=1}^n H \left(X_{{\lambda}+3i} \left | \right. F_{\mathcal{A} \backslash {\lambda}+3}^n,Y_{{\lambda}+3}^{i-1},F_{{\lambda}+3}^{i-1}\right) \right.
\\& \left.-
\sum_{i=1}^n H \left(X_{{\lambda}+3i} \left | \right. F_{\mathcal{A} \backslash {\lambda}+3}^n,Y_{{\lambda}+3}^{i-1},F_{{\lambda}+3}^{i-1},W_{\lambda}\right)
\right ]
\\ & \stackrel{{\rm{(b)}}}{\leq} \left( 1-\delta_{{\lambda}+3} \right) \left [n \right.
\\& \left. -\sum_{i=1}^n H \left(X_{{\lambda}+3i} \left | \right. F_{\mathcal{A} \backslash {\lambda}+3}^n,Y_{{\lambda}+3}^{i-1},F_{{\lambda}+3}^{i-1},W_{\lambda +3},Z_{{\lambda}+3}^{i-1}\right) \right ]
\\ & \stackrel{{\rm{(c)}}}{=}\left( 1-\delta_{{\lambda}+3} \right) n - \frac{nk_{{\lambda}+3}}{\delta_{{\lambda}+3\text{E}}},
\end{align*}
where: (i) the equality in ${\rm{(a)}}$ follows from the definition of mutual information; (ii) the inequality in ${\rm{(b)}}$ is because $H \left(X_{\lambda+3i} \right) \leq 1$ and because of the `conditioning reduces the entropy' principle; (iii) finally, the equality in ${\rm{(c)}}$ follows by using \eqref{eq:corr3}.
\\
{\bf{Sixth constraint.}}
From \eqref{eq:corr1} and \eqref{eq:corr2} we have $n e_3 + \frac{n k_3}{\delta_{3\text{E}}} = H \left( Y_3^n | W_{[1:2]},F_{\mathcal{A} }^n \right) $. 
By using this and Fano's inequality we have
\begin{align*}
&n \left(R_1 + R_2 \right)  + n e_3 + \frac{n k_3}{\delta_{3\text{E}}} & 
\\& \leq I \left( W_{[1:2]}; Y_{4}^n,Y_{5}^n, F_{\mathcal{A}}^n \right) + H \left( Y_3^n | W_{[1:2]},F_{\mathcal{A} }^n \right)  + n \epsilon
\\ & \stackrel{{\rm{(a)}}}{\leq} I \left( W_{[1:2]}; Y_3^n, F_{\mathcal{A}}^n \right) + H \left( Y_3^n | W_{[1:2]},F_{\mathcal{A} }^n \right) + n \epsilon
\\ & \stackrel{{\rm{(b)}}}{=} I \left( W_{[1:2]}; Y_3^n \left | \right. F_{\mathcal{A}}^n \right) + H \left( Y_3^n | W_{[1:2]},F_{\mathcal{A} }^n \right) + n \epsilon
\\ & \stackrel{{\rm{(c)}}}{=} H \left( Y_3^n \left | \right. F_{\mathcal{A}}^n \right)+ n \epsilon
\\ & \stackrel{{\rm{(d)}}}{\leq} n \left( 1-\delta_3 \right)+ n \epsilon,
\end{align*}
where: (i) the inequality in ${\rm{(a)}}$ follows because of the Markov chain $W_1,W_2-Y_3^n, F_{\mathcal{A}}^n-Y_{4}^n,Y_{5}^n$;
(ii) the equality in ${\rm{(b)}}$ follows because of the independence between $F_{\mathcal{A}}^n$ and $\left(W_1,W_2 \right)$; (iii) the equality in ${\rm{(c)}}$ follows from the definition of mutual information; (iv) finally, the inequality in ${\rm{(d)}}$ follows since node $\mathsf{M}_2$ receives at most $n \left( 1-\delta_3 \right)$ packets.
\\
{\bf{Seventh constraint.}} From \eqref{eq:corr4} and \eqref{eq:extraLemma} we have
\begin{align*}
& n \left ( \frac{k_1}{\delta_{1 \text{E}}} + \frac{k_2}{\delta_{2 \text{E}}} \right ) \frac{(1-\delta_3) \delta_{3 \text{E}}}{1-\delta_3 \delta_{3 \text{E}}}
\\& =
\frac{(1-\delta_3) \delta_{3 \text{E}}}{1-\delta_3 \delta_{3 \text{E}}} \left [H \left( Y_1^n | W_1,F_{\mathcal{A} }^n \right) + H \left( Y_2^n | W_2,F_{\mathcal{A} }^n \right)\right ]
\\& \stackrel{{\rm{(a)}}}{\geq}
\frac{(1-\delta_3) \delta_{3 \text{E}}}{1-\delta_3 \delta_{3 \text{E}}} \left [H \left( Y_1^n | W_{[1:2]},F_{\mathcal{A} }^n \right) \right.
\\& \left. + H \left( Y_2^n | W_{[1:2]},F_{\mathcal{A} }^n \right)\right ]
\\& \stackrel{{\rm{(b)}}}{\geq}
\frac{(1-\delta_3) \delta_{3 \text{E}}}{1-\delta_3 \delta_{3 \text{E}}} H \left( Y_1^n,Y_2^n| W_{[1:2]},F_{\mathcal{A} }^n \right)
\\& \stackrel{{\rm{(c)}}}{\geq} \frac{(1-\delta_3) \delta_{3 \text{E}}}{1-\delta_3 \delta_{3 \text{E}}} I \left( Y_1^n,Y_2^n; Z_3^n, Y_3^n| W_{[1:2]},F_{\mathcal{A} }^n \right)
\\& \stackrel{{\rm{(d)}}}{=}  \frac{(1-\delta_3) \delta_{3 \text{E}}}{1-\delta_3 \delta_{3 \text{E}}} \cdot
\\& \sum_{i=1}^n I \left( Y_1^n,Y_2^n; Z_{3i}, Y_{3i}| W_{[1:2]},F_{\mathcal{A} }^n, Z_3^{i-1},Y_3^{i-1} \right)
\\& = \frac{(1-\delta_3) \delta_{3 \text{E}}}{1-\delta_3 \delta_{3 \text{E}}} \left( 1-\delta_{3\text{E}} \delta_3 \right) \cdot
\\&\sum_{i=1}^n I \left( Y_1^n,Y_2^n; X_{3i}| W_{[1:2]},F_{\mathcal{A}\backslash 3}^n, Z_3^{i-1},Y_3^{i-1},F_3^{i-1},F_{3i}^n \right)
\\& \stackrel{{\rm{(e)}}}{=}  \frac{(1-\delta_3) \delta_{3 \text{E}}}{1-\delta_3 \delta_{3 \text{E}}} \left( 1-\delta_{3 \text{E}} \delta_3 \right) \cdot
\\& \sum_{i=1}^n H \left( X_{3i}| W_{[1:2]},F_{\mathcal{A}\backslash 3 }^n, Z_3^{i-1},Y_3^{i-1},F_3^{i-1} \right)
\\& \stackrel{{\rm{(f)}}}{=} n k_3 + n e_3 \frac{(1-\delta_3) \delta_{3\text{E}}}{1-\delta_3 \delta_{3\text{E}}},
\end{align*}
where: 
(i) the inequality in ${\rm{(a)}}$ is due to the `conditioning reduces the entropy' principle; (ii) the inequality in ${\rm{(b)}}$ is because $H \left( A,B\right) \leq H \left( A \right) + H \left( B \right)$;
(iii) the inequality in ${\rm{(c)}}$ follows since the entropy of a discrete random variable is positive; 
(iv) the equality in ${\rm{(d)}}$ is due to the chain rule of the mutual information; 
(v) the equality in ${\rm{(e)}}$ follows since $X_{3i}$ is uniquely determined given $\left( Y_1^n,Y_2^n, F_{\mathcal{A}}^n\right)$ and because of the Markov chain $X_{3i} -W_1,W_2,F_{\mathcal{A} \backslash 3}^n,Y_3^{i-1},Z_3^{i-1},F_3^{i-1} - F_{3i}^n$; (vi) finally, the equality in ${\rm{(f)}}$ follows from \eqref{eq:corr1} and \eqref{eq:corr2}.
\\
{\bf{Eighth constraint.}}
From \eqref{eq:corr1} and \eqref{eq:corr2} we have
\begin{align*}
& n \left (e_3 + \frac{k_3}{\delta_{3 \text{E}}} \right ) \frac{(1-\delta_{\kappa}) \delta_{\kappa \text{E}}}{1-\delta_{\kappa} \delta_{\kappa \text{E}}}
\\& =
\frac{(1-\delta_{\kappa}) \delta_{\kappa \text{E}}}{1-\delta_{\kappa} \delta_{\kappa \text{E}}} H \left( Y_3^n | W_{[1:2]},F_{\mathcal{A} }^n \right)
\\& \stackrel{{\rm{(a)}}}{\geq} \frac{(1-\delta_{\kappa}) \delta_{\kappa \text{E}}}{1-\delta_{\kappa} \delta_{\kappa \text{E}}} I \left( Y_3^n; Z_{\kappa}^n, Y_{\kappa}^n| W_{[1:2]},F_{\mathcal{A} }^n \right)
\\& \stackrel{{\rm{(b)}}}{=}  \frac{(1\!-\!\delta_{\kappa}) \delta_{\kappa \text{E}}}{1\!-\!\delta_{\kappa} \delta_{\kappa \text{E}}} \sum_{i=1}^n I \left( Y_3^n; Z_{\kappa i}, Y_{\kappa i}| W_{[1:2]},F_{\mathcal{A} }^n, Z_{\kappa}^{i-1},Y_{\kappa}^{i-1} \right)
\\& =  \frac{(1-\delta_{\kappa}) \delta_{\kappa \text{E}}}{1-\delta_{\kappa} \delta_{\kappa \text{E}}} \left( 1-\delta_{\kappa \text{E}} \delta_{\kappa} \right) \cdot
\\&\sum_{i=1}^n I \left( Y_3^n; X_{\kappa i}| W_{[1:2]},F_{\mathcal{A}\backslash \kappa}^n, Z_{\kappa}^{i-1},Y_{\kappa}^{i-1},F_{\kappa}^{i-1},F_{\kappa i}^n \right)
\\& \stackrel{{\rm{(c)}}}{=}  \frac{(1-\delta_{\kappa}) \delta_{\kappa \text{E}}}{1-\delta_{\kappa} \delta_{\kappa \text{E}}} \left( 1-\delta_{\kappa \text{E}} \delta_{\kappa} \right) \cdot
\\& \sum_{i=1}^n H \left( X_{\kappa i}| W_{[1:2]},F_{\mathcal{A}\backslash \kappa}^n, Z_{\kappa}^{i-1},Y_{\kappa}^{i-1},F_{\kappa}^{i-1},F_{\kappa i}^n \right)
\\& \stackrel{{\rm{(d)}}}{=}  \frac{(1-\delta_{\kappa}) \delta_{\kappa \text{E}}}{1-\delta_{\kappa} \delta_{\kappa \text{E}}} \left( 1-\delta_{\kappa \text{E}} \delta_{\kappa} \right) \cdot
\\&  \sum_{i=1}^n H \left( X_{\kappa i}| W_{[1:2]},F_{\mathcal{A}\backslash \kappa }^n, Z_{\kappa}^{i-1},Y_{\kappa}^{i-1},F_{\kappa}^{i-1} \right)
\\& \stackrel{{\rm{(e)}}}{=} n k_{\kappa},
\end{align*}
where: 
(i) the inequality in ${\rm{(a)}}$ follows since the entropy of a discrete random variable is positive; 
(ii) the equality in ${\rm{(b)}}$ is due to the chain rule of the mutual information; 
(iii) the equality in ${\rm{(c)}}$ follows since $X_{\kappa i}$ (with $\kappa \in [4:5]$) is uniquely determined given $\left( Y_3^n, F_{\mathcal{A}}^n\right)$; 
(iv) the equality in ${\rm{(d)}}$ follows because of the Markov chain $X_{\kappa i} -W_{[1:2]},F_{\mathcal{A} \backslash \kappa}^n,Y_{\kappa}^{i-1},Z_{\kappa}^{i-1},F_{\kappa}^{i-1} - F_{\kappa i}^n$; 
(v) finally, the equality in ${\rm{(e)}}$ follows from \eqref{eq:corr3}.

\bibliographystyle{IEEEtran}
\bibliography{isit2016}
\end{document}